\documentclass[acmsmall]{acmart}

\usepackage{booktabs} 
\usepackage{enumerate}
\usepackage{graphicx}
\usepackage{amsmath}
\usepackage{multirow}
\usepackage{footnote}
\usepackage{threeparttable}
\usepackage{amsmath}
\usepackage{textcomp}
\usepackage{lipsum}
\usepackage{hyperref}
\usepackage{nicefrac}
\usepackage{xcolor}

\newcommand{\mycolor}{black}

\newcommand{\figref}[1]{Fig.~\ref{#1}}

\newcommand{\secref}[1]{Section~\ref{#1}}
\newcommand{\tabref}[1]{Table~\ref{#1}}

\usepackage{natbib}

\usepackage[ruled,linesnumbered]{algorithm2e}
\usepackage{algpseudocode}

\usepackage{graphicx}

\usepackage{soul}

\usepackage{caption}
\usepackage{subcaption}

\AtBeginDocument{%
  \providecommand\BibTeX{{%
    \normalfont B\kern-0.5em{\scshape i\kern-0.25em b}\kern-0.8em\TeX}}}

\setcopyright{acmcopyright}
\copyrightyear{2023}
\acmYear{2023}

\acmJournal{JACM}
\acmArticle{1}
\acmMonth{5}

\begin{document}

\title{LL-GNN: Low Latency Graph Neural Networks on FPGAs for High Energy Physics}

\author{Zhiqiang Que}
\email{z.que@imperial.ac.uk}
\affiliation{%
  \institution{Department of Computing, Imperial College London}
  \country{UK}
}

\author{Hongxiang Fan}
\affiliation{
  \institution{Department of Computing, Imperial College London}
  \country{UK}
}


\author{Marcus Loo}
\affiliation{
  \institution{Department of Computing, Imperial College London}
  \country{UK}
}

\author{He Li}
\affiliation{
  \institution{School of Electronic and Engineering, Southeast University}
  \country{China}
}

\author{Michaela Blott}
\affiliation{
  \institution{AMD Adaptive and Embedded Computing Group (AECG) Labs}
  \country{Ireland}
}

\author{Maurizio Pierini}
\affiliation{
  \institution{European Organization for Nuclear Research (CERN)}
  \country{Switzerland}
}

\author{Alexander Tapper}
\affiliation{
  \institution{Department of Physics, Imperial College London}
  \country{UK}
}
\author{Wayne Luk}
\email{w.luk@imperial.ac.uk}
\affiliation{%
  \institution{Department of Computing, Imperial College London}
  \country{UK}
}

\renewcommand{\shortauthors}{Zhiqiang Que, et al.}

\begin{abstract}

This work presents a novel reconfigurable architecture for Low Latency Graph Neural Network (LL-GNN) designs for particle detectors, delivering unprecedented low latency performance. 
Incorporating FPGA-based GNNs into particle detectors presents a unique challenge since it requires sub-microsecond latency to deploy the networks for online event selection with a data rate of hundreds of terabytes per second in the Level-1 triggers at the CERN Large Hadron Collider experiments. 
This paper proposes a novel outer-product based matrix multiplication approach, which is enhanced by exploiting the structured adjacency matrix and a column-major data layout. In addition, we propose a custom code transformation for the matrix multiplication operations, which leverages the structured sparsity patterns and binary features of adjacency matrices to reduce latency and improve hardware efficiency. Moreover, a fusion step is introduced to further reduce the end-to-end design latency by eliminating unnecessary boundaries. Furthermore, a GNN-specific algorithm-hardware co-design approach is presented which not only finds a design with a much better latency but also finds a high accuracy design under given latency constraints. To facilitate this, a customizable template for this low latency GNN hardware architecture has been designed and open-sourced, which enables the generation of low-latency FPGA designs with efficient resource utilization using a high-level synthesis tool. Evaluation results show that our FPGA implementation is up to 9.0 times faster and \textcolor{\mycolor}{achieves up to 13.1 times higher power efficiency} than a GPU implementation. Compared to the previous FPGA implementations, this work achieves 6.51 to 16.7 times lower latency. Moreover, the latency of our FPGA design is sufficiently low to enable deployment of GNNs in a sub-microsecond, real-time collider trigger system, enabling it to benefit from improved accuracy. The proposed LL-GNN design advances the next generation of trigger systems by enabling sophisticated algorithms to process experimental data efficiently.


\end{abstract}



\maketitle


\section{Introduction}

Graph neural networks (GNNs) have shown remarkable successes in wide applications with graph-structured data, such as recommendation systems~\cite{ying2018graph, fan2019graph, yang2019aligraph}, molecule property prediction~\cite{fout2017protein, wu2018moleculenet}, and various High Energy Physics (HEP) applications, such as jet tagging~\cite{moreno2020jedi}, charged particle tracking~\cite{ju2021performance,elabd2021graph}, and calorimeter energy measurements~\cite{qasim2019learning}. In the domain of high energy physics, particularly in large-scale experimental setups such as CERN's Large Hadron Collider (LHC), 
the effectiveness of GNNs has been recognized. GNNs can extract and process complex patterns from high-dimensional data, which aligns well with the requirements of such environments where data from particle collisions are highly interconnected. However, incorporation of GNNs into real-time data processing in these settings presents significant challenges, primarily due to the enormous data volume and strict low latency requirements.

Real-time data processing from high-energy proton collisions at the LHC is challenging because the particle detectors around the LHC ring produce hundreds of terabytes of data per second~\cite{coelho2021automatic, duarte2018fast} from collisions happening every 25 ns. In the next phase of the LHC, the upgraded High-Luminosity LHC (HL-LHC) experiments expect an explosion of data due to cutting edge detectors with improved resolution and increased area as well as volume. The large data volumes produced from the detectors are reduced by an embedded real-time processing system, known as the trigger, which keeps interesting collision events while discarding the others. The Level-1 Trigger~(L1T), which uses only FPGAs, requires sub-microsecond application processing latency in the LHC~\cite{CERN2020L1T}. If algorithm latency exceeds this strict limit, valuable data and significant events are lost.

This inherent requirement for ultra-low latency, combined with the computational intensity of GNNs, presents a unique challenge in the field of high energy physics. Indeed, GNNs require large amounts of computation and suffer from irregular memory accesses, leading to significant inference latency. This hinders their effective deployment in real-time, ultra-low latency applications like the L1T system in an LHC experiment as suggested by Moreno et al.~\cite{moreno2020jedi}. 
Accelerating GNN inference using reconfigurable accelerators such as FPGAs is essential in the LHC as it enables sophisticated processing in real-time with superior accuracy. Nonetheless, most existing FPGA-based GNN accelerators employ a single-engine architecture with fixed hardware to sequentially process layers or sub-layers (blocks) like GPUs, and the networks are processed in a recurrent fashion~\cite{yan2020hygcn, liang2020engn, zhang2021boostgcn, lin2021gcn, lin2022hp, kang2022grow}. However, they are not efficient for GNN execution when handling small graphs that necessitate ultra-low latency and high throughput, such as in scientific applications for particle identification. Furthermore, none of them is designed to operate under a hard latency constraint of one microsecond ($1\mu$s).

To address the gap, this work proposes a domain-specific Low Latency (LL)-GNN hardware architecture based on a layer-wise tailor-made pipeline to accelerate the GNNs for particle detectors. The layer-wise architecture has been used to speed up CNNs~\cite{blott2018finn, shen2017maximizing, zhang2020dnnexplorer} and RNNs~\cite{que2021accelerating, que2023reconfigurable}, but there remains a lack of research focused on accelerating GNNs. 
This work uses the GNN-based JEDI-net algorithm~\cite{moreno2020jedi} as an end-to-end application. JEDI-net delivers state-of-the-art accuracy for particle identification and has sparked a growing interest in the use of GNNs in trigger systems.
Our work introduces an outer-product based matrix-matrix multiplication (MMM) approach for calculating the GNN aggregation function, which is further enhanced by exploiting the structured adjacency matrix. In addition, the proposed architecture adopts a column-major order instead of conventional row-major order for the data layout of intermediate results traversed along the hardware datapath, which avoids irregular memory access for intermediate results in GNNs. 
Moreover, we present a custom code transformation for MMMs of GNN feature transformation based on the character of interaction-network based GNNs that utilize a fully connected graph as input. This approach avoids the costly matrix multiplications between the adjacency matrix and the input feature matrix, thereby enhancing computation efficiency. 

Furthermore, instead of running GNNs in a coarse-grained pipeline as shown in our prior work~\cite{que2022aicas, que2022optimizing}, we propose fusion of sub-layers as much as possible, transforming multiple coarse-grained pipeline stages into a single stage with a fine-grained pipeline inside. 
The fusion eliminates unnecessary handshake acknowledgments and dual buffers between coarse-grained pipeline stages, effectively reducing end-to-end latency. 
To further optimize latency, a Finite State Machine (FSM)-based code structure is proposed to handle imperfect loops that may emerge after fusion.
Finally, we present a GNN-specific algorithm-hardware co-design approach to optimize the algorithm and hardware simultaneously, exploring the optimal configuration to enhance overall performance. Our LL-GNN design successfully achieves sub-microsecond latency, which makes the algorithm compatible with the HL-LHC conditions where there is a hard ultra-low latency constraint, while also enabling improved accuracy.

To the best of our knowledge, this is the first FPGA-based design of GNNs with sub-microsecond latency for particle identification for the detectors at the CERN HL-LHC experiments. This work would help to improve next-generation trigger systems, enabling powerful algorithms to process massive LHC data effectively. 

We make the following contributions in this paper:
\begin{itemize}




\item A low latency layer-wise hardware architecture for GNNs with many novel design optimizations, including outer-product based matrix multiplication, column-major order, custom code transformation and the fusion of sub-layers as well as a GNN-specific algorithm and hardware co-optimization approach, resulting in sub-microsecond latency and high hardware efficiency.  






\item A scalable, efficient open-source template\footnote{https://github.com/walkieq/GNN-JEDInet-FPGA} for GNN-based JEDI-net, which enables the generation of low-latency FPGA designs with efficient resource utilization leveraging high-level synthesis (HLS) tools. 

\item A comprehensive evaluation of the proposed method and hardware architecture.

\end{itemize}


Although the GNN coefficients are HEP specific, the proposed optimizations and techniques have broader implications for other GNN applications beyond particle identification. 
\textcolor{\mycolor}{Most of the techniques and optimizations introduced in Section~\ref{sec:design_and_optimization} and Section~\ref{sec:2level_parallelism}$\sim$\ref{sec:co-design} are general} and can be beneficial to a range of GNN applications, not just limited to the JEDI-net. These optimizations include strategies for streamlining matrix operations, hardware pipelining, and ways to optimize FPGA resources for maximum performance, minimizing design latency and maximizing acceleration efficiency.
Moreover, the techniques devised for low latency optimization can benefit numerous other applications on FPGAs, especially those that require real-time processing. 





\subsubsection*{Relationship to Prior Publications}
This paper expands on our previous work~\cite{que2022aicas, que2022optimizing} which primarily focuses on hardware optimizations and targets low initiation interval, but faces challenges with high latency. This work addresses two limitations of the prior work. First, we focus on optimizing end-to-end latency while prior work concerns with initiation interval. Our new approach fuses together as many coarse-grained pipeline stages as possible into a single fine-grained pipeline stage, resulting in low latency designs. Second, we cover co-optimization of both algorithm and hardware, in contrast to the prior work that only focuses on hardware optimizations. This work explores the design performance trade-off under both user-defined algorithm and hardware constraints, such as achieving the highest accuracy with a latency requirement of less than 1$\mu$s on a given FPGA. The optimized design provides a significant reduction in end-to-end latency, achieving speedups of 6.51 to 16.7 times along with enhanced model accuracy when compared to our previous studies~\cite{que2022aicas, que2022optimizing}. For the first time, our novel architecture supports GNNs of JEDI-net to operate in less than 1$\mu$s on FPGAs. To facilitate the deployment of these optimizations, this paper provides an open-source template for generating low-latency FPGA implementations of GNN-based JEDI-net with efficient resource utilizations, which has not been covered by previous work.

\section{Background}


\subsection{Graph Neural Network and Interaction Network}
GNNs can adapt their graph-based structure to an input graph with an iterative process of information aggregation across nodes to learn the complex dependencies of a system. In essence, GNNs create new graph embedding and leverage these for making predictions at edge-level, node-level, or graph-level. The interaction network~\cite{battaglia2016interaction} is a powerful graph based framework for reasoning about objects and relations in complex and dynamic systems. The network takes a graph as input, where nodes represent objects and edges represent interactions between them. Each node and edge has an associated set of features that describe the state of the object or interaction. It learns to capture complex interactions that can be used to predict future states and abstract physical properties. 

\begin{figure} 
   \centering
  \subfloat[]{
   \includegraphics[width=0.38\linewidth]{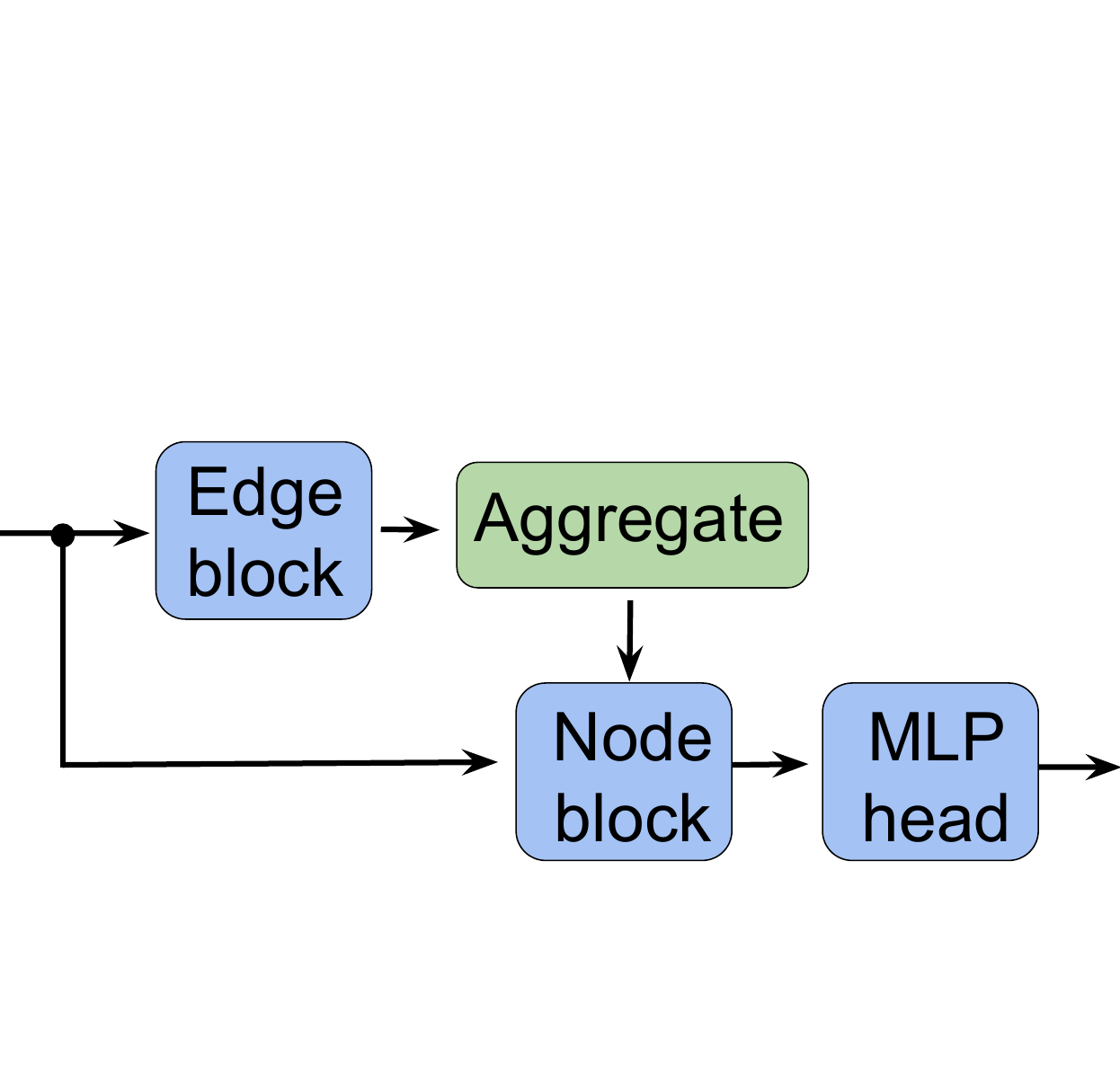}}
    \hspace*{\fill}
  \subfloat[]{%
    \includegraphics[width=0.6\linewidth]{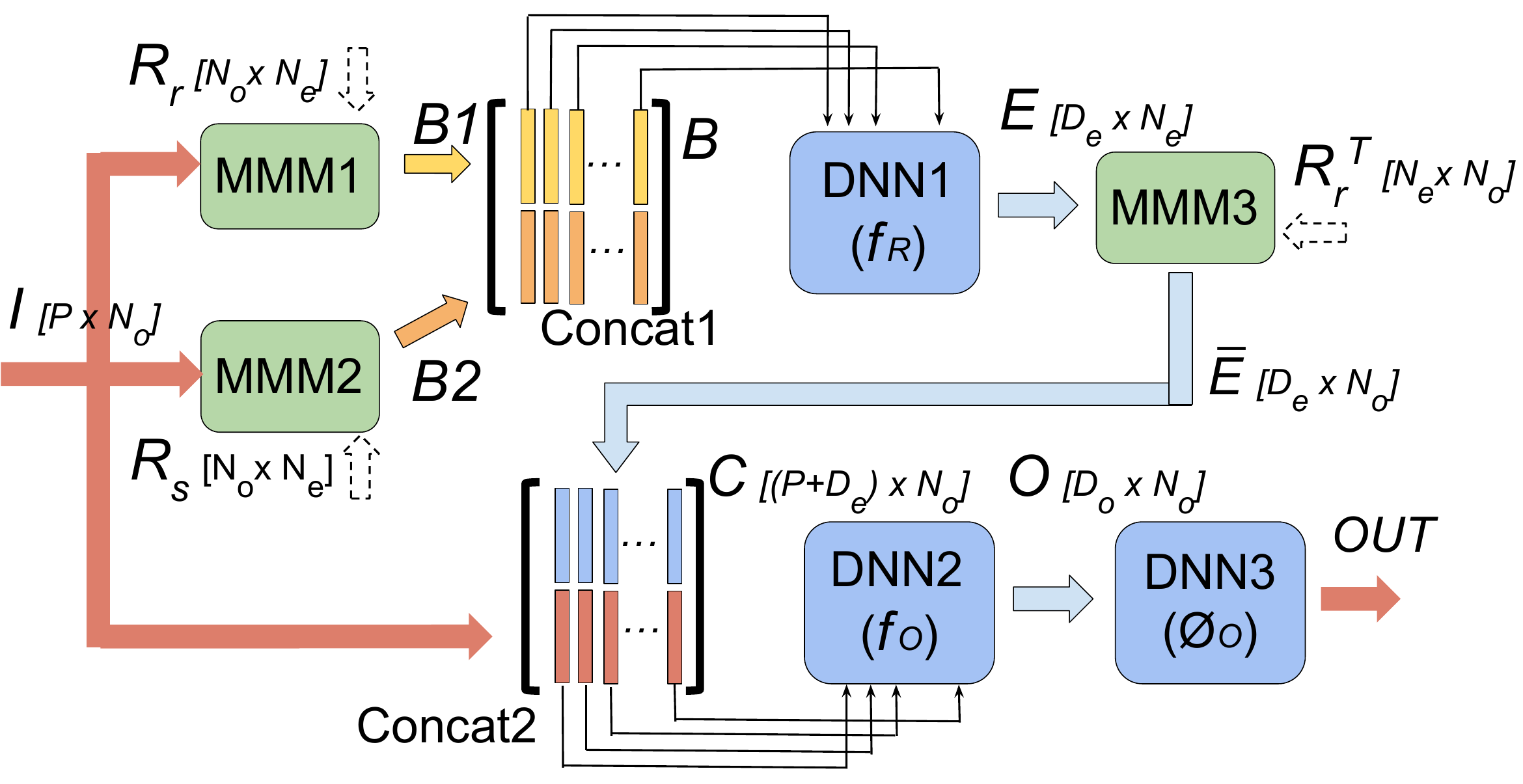}}
  \caption{ (a) An example of interaction network-based GNN with edge block, aggregation, node block and MLP-head. (b) Overview of the JEDI-net architecture.)}
  \label{fig:sys} 
\end{figure}

An interaction-network based GNN processes the input graph in three main steps in each iteration. First, it applies an edge function to each edge in the graph, taking as input the features of the edge and the features of the nodes it connects. This edge function models the interaction between the objects, producing an updated edge feature (often called a "message").
Second, these messages are consolidated by an aggregation function over the nodes interconnected to each edge, providing a summary of the incoming messages. The process of producing and consolidating messages is known as a message-passing operation. The aggregation function, here assumed to be a summation, transposes the edge-specific information to node-specific outputs by gathering information based on the connected node indices.
Third, a node function is then applied to each node, taking as input the node's original features and the aggregated messages. This node function models how the object's state is affected by its interactions, producing an updated node feature. An additional GNN-head function gives the final graph-level prediction. An example full forward pass, comprised of edge and node blocks used to conduct a graph-level prediction, is shown in Fig.~\ref{fig:sys}(a).




\begin{figure}
\begin{center}
\includegraphics[width=0.7\linewidth]{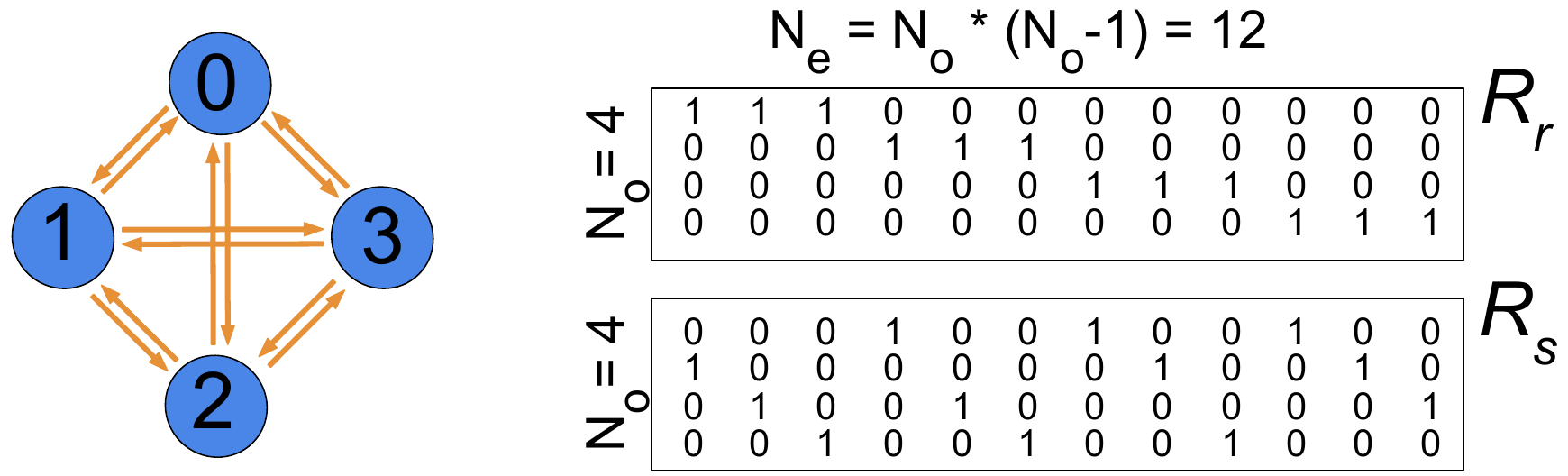}
\end{center}
   \caption{An example of an interaction-network based fully-connected graph with 4 nodes and the corresponding 12 uni-directional edges (left) with its receiving matrix $R_r$ as well as the sending matrix $R_s$ (right). 
   }
   \vspace{0.2cm}
\label{fig:adjacency}
\end{figure}

\subsection{JEDI-net for Particle Identification}\label{subsec:jedi-net}
Early trigger methodologies~\cite{cheatham2016atlas} relied solely on jet energy to make a decision. Recently, machine learning based strategies are being used for their superior accuracy. However, due to the low latency requirement, only simple Multi-Layer Perceptron (MLP) networks~\cite{coelho2021automatic, duarte2018fast} on FPGAs are introduced. High accuracy in the trigger is crucial to keep only the most interesting events while keeping the output bandwidth low~\cite{coelho2021automatic}. As a solution, the study in~\cite{moreno2020jedi} presents JEDI-net, a Graph Neural Network (GNN) based algorithm, that delivers state-of-the-art accuracy for particle identification. This development has led to a surge in the demand for GNNs in trigger systems.

JEDI-net~\cite{moreno2020jedi} is a fully connected GNN based on the interaction network architecture, which identifies the particles produced at the LHC by analyzing jets. A jet is a narrow cone of hadrons and other particles produced by the hadronization of a particle. 
The input of JEDI-net can be represented as a graph, $G=\langle I,R \rangle$ with the nodes, $I$, corresponding to physics particles, and the edges, $R$, to the relations. It is a fully connected GNN. The input of nodes ($I$) is defined as a $P \times N_o$ matrix, whose columns represent the node’s $P$-length feature vectors, and $N_o$ is the number of particles in a jet. The relations  are represented by a triplet, $R=\langle R_r,R_s,{R_r}^T \rangle$, where $R_r$ and $R_s$ are $N_o \times N_E$ binary matrices which index the receiver and sender nodes, respectively. 
Each column of $R_r$ and $R_s$ is a one-hot vector and it indicates the receiver node’s index; $R_s$ indicates the sender similarly.
The number of edges, $N_e$, is $N_o \times (N_o - 1)$ since the input graph is fully connected with directional edges. 

Fig.~\ref{fig:sys}(b) shows an overview of JEDI-net. In addition, \figref{fig:adjacency} shows the $R_r$ and $R_s$ of this fully connected GNN for an example.
To illustrate the idea, the number of particles (nodes) is 4 in this example. But please note a real case will have more particles. 
The input $I$ matrix is multiplied by the $R_r$ and $R_s$ matrices and the results are then concatenated to form a $B$ matrix, having dimension $2P \times N_e$. Each column of the $B$ matrix represents an edge, i.e. a particle-to-particle interaction. The $2P$ elements of each column are the features of the sending and receiving nodes for that edge. 
A trainable deep neural network (DNN) function $f_{R}:\mathbb R^{2P} \rightarrow \mathbb R^{D_e}$ is then applied to each column of $B$ and produces a matrix $E$. Then $\Bar{E}=ER_r^T$ is computed in MMM3 (see Fig.~\ref{fig:sys}(b)) to gather the cumulative effects of interactions received by a given node. 
Thus, the cumulative effects of the interactions at a given node are obtained by summing the $D_e$ hidden features over the incoming edges, which is implemented by computing $\Bar{E}=ER_r^T$ in the MMM3 unit. 
The $\Bar{E}$ and $I$ are then concatenated to form the $C$ matrix, working as a shortcut connection. Each column of the $C$ matrix represents a constituent in the jet, which is expressed as a $(P+D_e)$-dimensional feature vector, containing $P$ input features and $D_e$ hidden features. It represents the combined effect of all the interactions between particles. 
Another trainable function $f_{O}$ is presented to build a post-interaction representation of each jet constituent. It is performed on each column of $C$ to produce the $O$ matrix, having dimension $D_o \times N_o$. A final trainable function $\phi_{O}$ returns the probability for that jet to belong to each of the five categories. $f_{R}$, $f_{O}$ and $\phi_{O}$ are implemented as MLPs.



\section{Design and Optimization}~\label{sec:design_and_optimization}
This section introduces the hardware architecture and multiple optimizations to accelerate the interaction network based GNNs. We adopt a 3-step optimization strategy, named "divide, conquer and fusion", to achieve low latency for GNN inferences. During the "divide" and "conquer" steps, we split GNNs into multiple sub-layers and perform dedicated optimizations, such as outer-product based MMM (Section~\ref{sec:outer_mmm}) approach and leveraging the structured sparsity of adjacency matrices (Section~\ref{sec:sparsity}) for GNNs. 
Then, in the "fusion" step (Section~\ref{sec:fuse}), multiple sub-layers are fused together to eliminate boundaries and buffers between different pipeline stages, resulting in low latency and high hardware efficiency. 

\subsection{Outer-product Based Matrix Multiplication}~\label{sec:outer_mmm}
\begin{figure} 
\centering
    \subfloat[The calculation of the first vector of the resultant matrix.\label{fig:outer1}]{
    \includegraphics[width=0.7\linewidth]{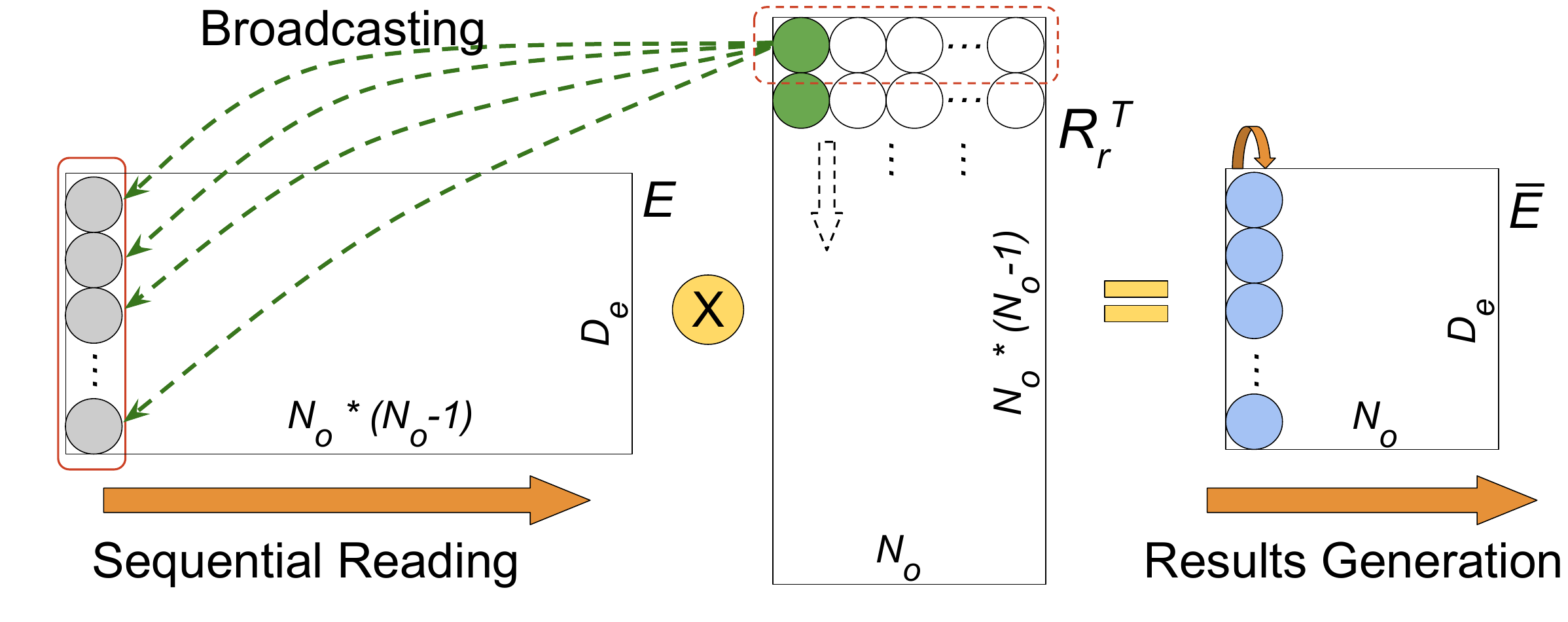} }
  \\
    \subfloat[The calculation of the second vector.\label{fig:outer2}]{
    \includegraphics[width=0.7\linewidth]{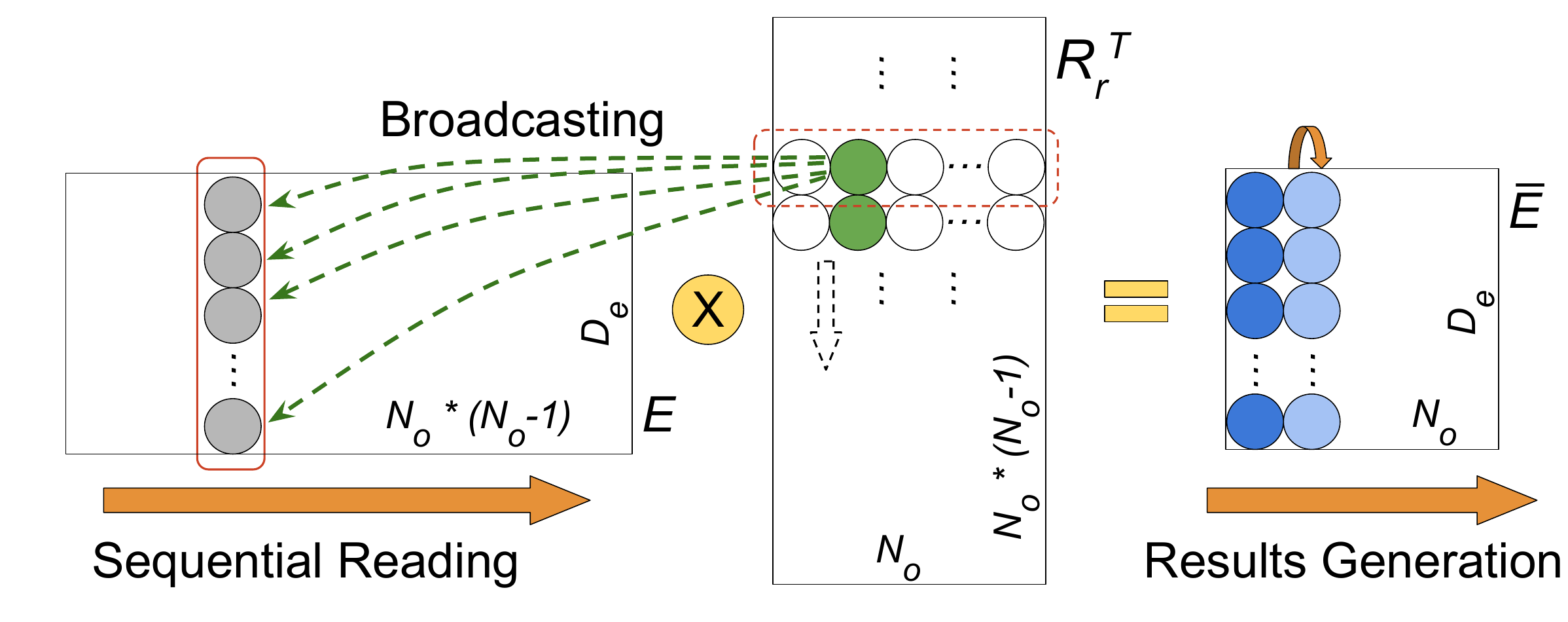} }
  \caption{Outer-product based matrix multiplication with column-major order and structured sparsity.}

  \label{fig:outer} 
\end{figure}

Matrix multiplication is a fundamental operation in the calculation of GNNs. It typically involves the inner-product of rows from the first matrix and columns from the second matrix. For instance, to compute the GNN aggregate function $\Bar{E}=ER_r^T$ (MMM3), as shown in Fig.~\ref{fig:sys}(a) and (b), one needs an entire row of the $E$ matrix and a full column from the $R_r^T$ matrix to perform the inner-product for each entry of $\Bar{E}$. 
However, as the $E$ matrix is generated column by column, this introduces a long latency when using the inner-product based MMM since it must wait until an entire row of the $E$ matrix is ready, leading to a considerable delay. To address this issue, this work proposes an outer-product based matrix multiplication to process the GNN aggregate function (MMM3).
Instead of using a full row from $E$ matrix, now a full column from the $E$ matrix is multiplied by one element from $R_r^T$ matrix to generate the partial result of the first column of the result matrix $\Bar{E}$ as shown in Fig.~\ref{fig:outer}(a). This partial result is then accumulated to form a complete column of the result matrix. Given that the $E$ matrix is produced column by column, MMM3 can start as soon as the first column of the $E$ is ready. To efficiently support the outer-product based MMM, the column-major order data layout (discussed in Section~\ref{sec:cm}) is used for representing the intermediate results (i.e., 2D matrix arrays). Thus, input elements can be grouped as a vector (i.e., an entire column as a vector) and can be processed efficiently with high parallelism because the data can be fetched sequentially. It substantially minimizes the waiting time for the aggregate function in GNNs, thus reducing the overall design latency. 



\begin{algorithm}[t]
  \caption{The pseudocode of the outer-product based MMM.}
  \label{algo:mmm3}
  \LinesNumbered
 \begin{footnotesize} 
  \SetKwFunction{FMain}{MMM3}
  \SetKwProg{Fn}{Function}{:}{}
  \Fn{\FMain{ $E$, $\Bar{E}$ }}{
  // The layout of matrices is column-wise \\
      \For{$i = 0$ to $N_o$}{
        \For{$k = 0$ to $N_o-1$} {
            \For{$j = 0$ to $D_e$}{
                $index=i \times (N_o-1) +k$; \\
                $tmp = (k==0)\ \ ?\ \ 0 : acc[j]$; \\
                $acc[j] = tmp + E[index][j]$; \\
                
            }
        }
        \For{$m = 0$ to $D_E$}{
            $\Bar{E}[i][m] = acc [m]$;
        }
    }
  }
  \textbf{End Function}
\end{footnotesize}
  
\end{algorithm}

In addition, the custom code transformation (discussed in Section~\ref{sec:sparsity}), can be adapted to enhance the proposed outer-product based MMM, which exploits the sparsity patterns inherent in the $R_r^T$ adjacency matrix and the binary features.
It can avoid costly multiplications and involves only load and store operations with a small number of additions for this MMM unit.
Detailed pseudocode of the enhanced outer-product based MMM is presented in Algorithm~\ref{algo:mmm3}. Generally, in GNNs, the aggregation has a large amount of data access and relatively small amount of computation, resulting in low hardware efficiency~\cite{zhong2023cognn}. However, our approach takes advantage of the sparse structures and binary features in the adjacency matrix. 
The input $R_r$ matrix is binary and each column is one-hot as explained in Section~\ref{subsec:jedi-net}. Therefore, the multiplication operations are unnecessary since $R_r^T$ is binary. Besides, only $\frac{1}{N_o}$ of the total number of additions are required. 
Moreover, due to the structured pattern of the adjacency matrix, the value and pattern of adjacency matrices can be incorporated into the loop index to avoid irregular memory access.


One limitation of the outer-product based MMM is that it produces a full size resultant matrix with only partial results until the end~\cite{que2021recurrent}. This process requires substantial memory write bandwidth. The AWB-GCN~\cite{geng2020awb} presents a column-wise product architecture, a variant of outer-product MMM, which involves one column from the first matrix and only one element from the second matrix, such that the output only affects one resultant vector. In our design, each row is one-hot, which means that only one vector in the resultant matrix is valid at a time. Moreover, we further exploit the structured pattern of $R_r$ so that the partial results only update the corresponding vectors in the resultant matrix sequentially, preventing irregular memory writes and thereby reducing latency. Furthermore, due to the structured pattern in $R_r$, the proposed approach only requires the $E$ matrix to be read once, which also reduces the bandwidth of memory access. 
Fig.~\ref{fig:outer}(b) shows how the design reads the vectors from column $(N_o-1)$ to $2*(N_o-1)$ in the $E$ matrix and generates the second vector in the resultant matrix. 

The latency could be further reduced by involving multiple columns of input matrix. However, this requires multiple preceding hardware units to generate the equivalent number of columns during each cycle, thereby increasing hardware resource usage. 
Our proposed methods not only eliminate the costly matrix multiplication operations and reduce iterations, but also improve memory access efficiency by removing the necessity for adjacency matrix inputs. These techniques thus decrease design latency, and boost throughput as well as hardware efficiency.




\subsection{Column-major Order}\label{sec:cm}

\begin{figure} 
  \hspace*{\fill}
    \includegraphics[width=0.25\linewidth]{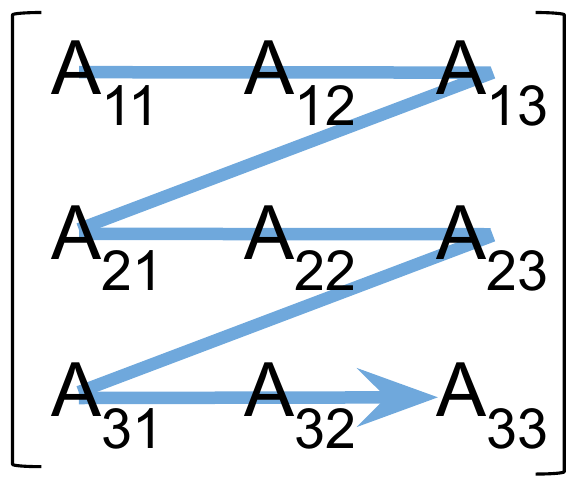}
  \hfill
    \includegraphics[width=0.25\linewidth]{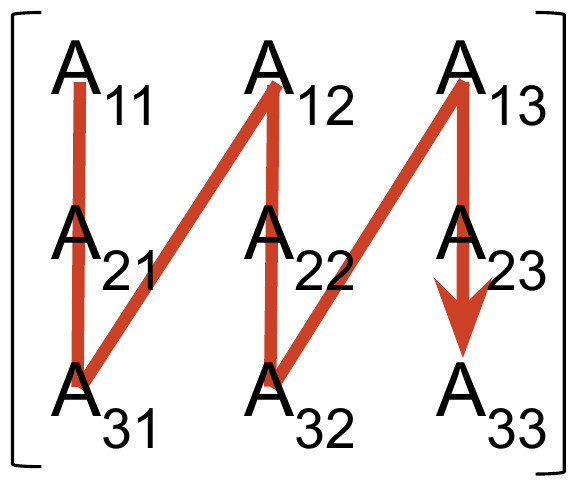}
  \hspace*{\fill}
  \caption{Row-major (left) and column-major (right) orders.}
  \label{fig:col_major_order} 
\end{figure}

The intermediate results in the layer-wise GNN hardware architecture are captured using two-dimensional (2D) arrays representing a matrix as shown in~\figref{fig:sys}(b). Row-major and column-major orders (\figref{fig:col_major_order}) are two data layout methods, which are critical for correctly passing arrays between hardware units. More importantly, choosing the right data layout method is also critical for hardware performance, which is the focus of this work. The difference between the two orders lies in which elements of an array are contiguous in memory. An appropriate data layout will have a significant impact on hardware performance.
When mapping 2D data onto a one-dimensional (1D) structure (i.e. memory) using a high level synthesis tool (e.g., Xilinx Vivado / Vitis HLS), often the default data layout is row-major order, which follows a C language legacy. 

However, row-major order for GNN-based JEDI-net will lead to poor spatial locality and hinder parallelism, since the functions $f_{R}$ and $f_{O}$ are applied to each column of the input matrix, as shown in~\figref{fig:sys}(b). The concatenation of two input matrices into one output matrix also works on columns, such as concatenating $B1$ and $B2$ to the $B$ matrix. 
With a row-major order data layout, the input data of these functions do not sit in memory contiguously so it is very time-consuming to fetch all the elements in a column. 
However, if the data are represented using a column-major order, in which the consecutive elements of a column reside next to each other, iterating over columns becomes easy because the data are accessed sequentially. 
Thus, this work proposes the column-major order to increase the data spatial locality for accelerating layer-wise GNNs efficiently, leading to good hardware performance. 


\begin{algorithm}[t]
  \caption{The pseudocode of the custom MMMs.}
  \label{algo:mmm}
  \LinesNumbered

\begin{footnotesize}
  \SetKwFunction{FMain}{MMM\_$B$}
  \SetKwProg{Fn}{Function}{:}{}
  \Fn{\FMain{ $I$, $B1$, $B2$ }}{
  // The layout of matrices is column-wise \\  
      \For{$i = 0$ to $N_O$}{
        \For{$k = 0$ to $N_O-1$}{ 

            // Reduced from $N_o \times (N_o-1)$ to $N_o-1$ \\
          
            \For{$j = 0$ to $P$}{
                $B1[k+i*(N_O-1)][j]=I[i][j]$;  \Comment{MMM1}
                \vspace{0.2cm}
                
                $index=(k < i)\ ?\ k:(k+1)$;  \Comment{MMM2}
                
                $B2[k+i*(N_O-1)][j]=I[index][j]$;\\ 
                \vspace{0.2cm}
                // Multiplications are avoided \\
                // Access of $R_r$ and $R_s$ matrix is avoided 
            }
        }
    }
  }
  \textbf{End Function}
\end{footnotesize}
  
\end{algorithm}

\subsection{Custom MMMs for GNN Feature Transformation}\label{sec:sparsity}

The GNN design detailed in~\cite{moreno2020jedi} relies on dense MMM operations to calculate the MMMs between the input feature matrix and adjacency matrices, specifically, $B1=IR_r$ and $B2=IR_s$. 
However, this approach proves to be resource-intensive and time-consuming. 
Interestingly, we observe that a significant proportion of the computational operations in these MMMs, particularly in fully-connected GNNs based on interaction networks, are redundant by leveraging the sparse structures and binary features within the adjacency matrices. 
In the case of interaction network-based GNNs with fully-connected graphs, every column within the receiving and sending matrices is one-hot. In addition, these matrices not only feature a binary structure but also have static patterns, as shown in~\figref{fig:adjacency}. 
The element ${(R_r)_{ij}}$ is set to 1 when the $i^{th}$ node receives the $j^{th}$ edge, and is 0 otherwise. Similarly, the element ${(R_s)_{ij}}$ is set to 1 when the $i^{th}$ node sends the $j^{th}$ edge and is 0 otherwise. 
Due to these unique characteristics, many computational operations are unnecessary. First, the multiplication operations are unnecessary because the $R_r$ and $R_s$ matrices only have binary values. Second, accumulation (addition) operations can be eliminated when doing the inner product of row vectors from the input feature $I$ matrix and column vectors from $R_r$ (or $R_s$). This is because each column of $R_r$ and $R_s$ is one-hot, resulting in only one product being produced. As a result, the iteration factor is reduced from $N_o \times (N_o-1)$ to $N_o-1$. This simplifies the calculation of $B1=IR_r$ and $B2=IR_s$ to merely load and store operations.
A detailed pseudocode of the custom code transformation for MMM1/2 is provided in Algorithm~\ref{algo:mmm}.

One challenges when dealing with GNNs is handling the irregular memory access of sparse adjacency matrices. Our approach can avoid such memory access by building the known patterns of sparsity directly into the loop index. 
In addition, the access of the input feature matrix for $B1$ and $B2$ is sequential in our designs since we adopt a column-major order data format as discussed in Section~\ref{sec:cm}. The proposed approach not only eliminates the costly MMM operations to increase the computational efficiency, but also avoids the irregular access of the adjacency matrices, which greatly reduces the design latency.

Although the patterns presented here are specific to fully-connected GNNs based on interaction networks, the idea of exploiting the adjacency matrix pattern is highly adaptable. We believe that the method of code transformation we've employed could be adjusted to improve other GNN networks that have a useful pattern of sparsity, leading to low latency designs.






\subsection{A Dataflow Architecture with Task-level Parallelism }\label{sec:tlp}
\begin{figure}
\begin{center}
\includegraphics[width=0.8\linewidth]{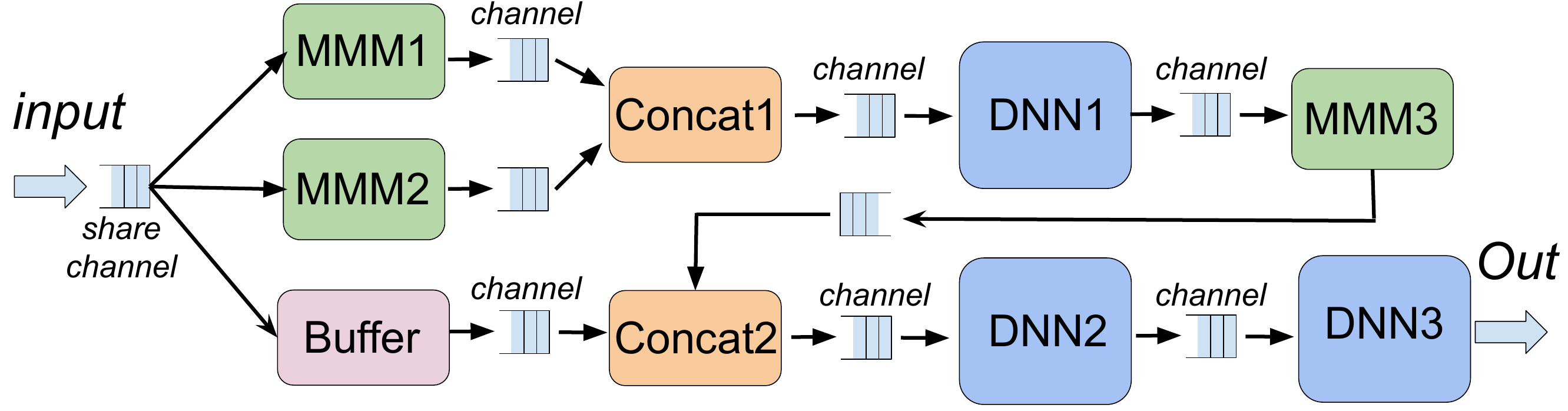}
\end{center}
   \caption{The dataflow of JEDI-net.}
\label{fig:dataflow}
\end{figure}

To improve performance and reduce latency, our previous designs~\cite{que2022aicas, que2022optimizing} partition the entire network into multiple sub-layers and utilize a dataflow architecture with task-level parallelism~\cite{blott2018finn, zhang2020dnnexplorer}. All the sub-layers are mapped on-chip, with each one as a task within a coarse-grained pipeline, allowing simultaneous execution. These tasks run as soon as their inputs become valid, thereby enabling efficient data flow between them. Ping-pong buffers are used as the channels between various tasks. Signals indicating when the buffers are full or empty are used for handshaking. 
\figref{fig:dataflow} shows the sub-layers (tasks) within in the GNN network, with blocks representing task operations and arrows representing data flow between tasks. The partitioning rule is based on the primary operations within the GNN network, as presented in Fig.~\ref{fig:sys}.

In addition, we implement dedicated optimizations, categorizing them into two classes: intra-task and inter-task optimizations. 
Given that each task operates independently, dedicated intra-task optimizations can be applied to each task, such as exploiting outer-product based MMM (discussed in Section~\ref{sec:outer_mmm}) and structured sparsity of adjacency matrices (discussed in Section~\ref{sec:sparsity}). Moreover, inter-task optimizations are also introduced to refine multiple tasks and their interactions, such as column-major order representation (discussed in Section~\ref{sec:cm}) to increase the spatial locality of intermediate results transferred between tasks, and initiation interval balancing~\cite{que2021accelerating} for optimizing the overall initiation interval across multiple tasks\textcolor{\mycolor}{, and a fusion technique  (discussed in Section~\ref{sec:fuse}) which fuses multiple tasks to remove extra boundaries, reducing end to end latency. The proposed GNN layered structure and coefficients are CERN LHC specific, while the dataflow architecture with task-level parallelism and the optimizations, especially the inter-task ones, can be adapted to other GNN networks or even to other types of neural networks. }

The initiation interval is decided by the largest initiation interval among all the units on the datapath. It is usually not necessary to unroll every unit in order to achieve the smallest initiation interval. Some hardware resources can be saved from the units that do not require full unrolling, and then these hardware resources can be reassigned to the bottleneck unit that dominates the whole design to reduce the initiation interval and latency~\cite{que2021accelerating} by an appropriate partitioning of hardware resources. While partitioning FPGA resources to enhance latency and initiation interval in a layer-wise architecture has been studied for CNNs~\cite{shen2017maximizing, zhang2020dnnexplorer, zhang2018dnnbuilder, gong2018maloc} and RNNs~\cite{que2021accelerating}, there is little work focusing on GNNs. This work addresses this gap by balancing the initiation intervals of the sub-layer units in GNNs by appropriate FPGA resources partitioning. 

\begin{figure} 
    \subfloat[]{%
    \includegraphics[width=0.55\linewidth]{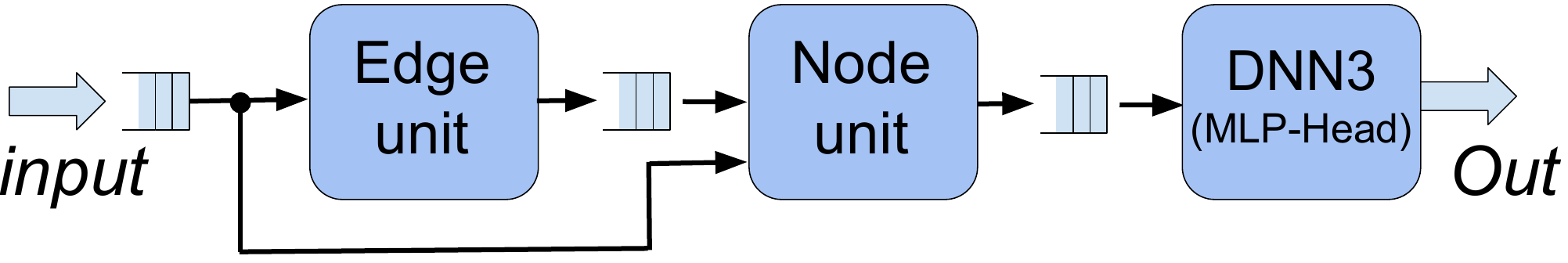}}
  \hfill
    \subfloat[]{%
    \includegraphics[width=0.40\linewidth]{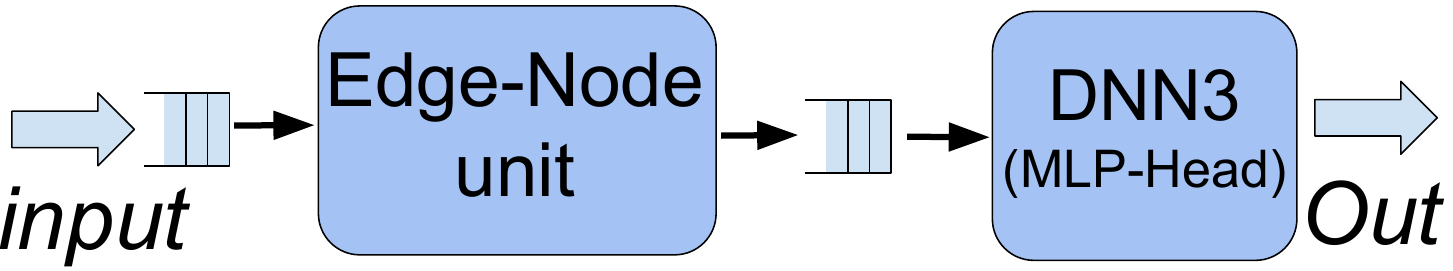}}
  \caption{(a) Fusion with separated edge and node functions. (b) Fusion with combined edge and node functions. }
  \vspace{0.2cm}
  \label{fig:dataflow_fusion} 
\end{figure}

\subsection{Divide, Conquer and Fuse}~\label{sec:fuse}
In the previous subsections, we "divide" the GNN model into multiple sub-layers and "conquer" them with tailed-made optimizations. Our previous designs~\cite{que2022aicas, que2022optimizing} deploy these sub-layers in a coarse-grained pipeline, achieving a low initiation interval. Nevertheless, the end-to-end latency remains higher than the $1\mu$s which is required for the CERN HL-LHC. To address this, we introduce a succeeding step called "fuse" which combines multiple optimized sub-layers into a single unit and runs it in a fine-grained pipeline to reduce overall design latency. 

Within GNNs, operations are typically performed repeatedly on graph nodes or edges. This implies that many processing units (or loops) share a common loop bound, that is, the number of nodes (or edges) in a graph, which makes many sequential units (loops) mergeable. Particularly, in our GNN designs, the matrix multiplication units are simplified after exploiting structure sparsity. Running them as individual stages within a coarse-grained pipeline not only leads to lots of wasted hardware resources but also increased latency due to the extra acknowledgments, including ping-pong buffers that exist between stages in the pipeline. As multiple sequential loops share the same loop bound, these loops can be merged into one to reduce the latency. 

For example, the MMM1/2, Concat1 and DNN1 units all have a loop bound of $N_e$ (number of edges), as they perform calculations on the edges of the graph. They can be consolidated into one large loop, which can then run in a fine-grained pipeline to reduce overall latency, as merging loops allows the logic within the loops to be optimized together. The Concat2 and DNN2 units, on the other hand, have a loop bound of $N_o$ (number of nodes), as they perform calculations on the nodes of the graph. An interesting aspect to consider is the aggregate unit (MMM3) which accumulates the effects of interactions from edges to nodes. This work proposes to divide this unit into two parts. The first part, covering lines 3 to 10 in Algorithm~\ref{algo:mmm3}, has a loop bound of $N_e$ and can be fused into a new unit called Edge unit, while the second part (lines 11 to 13 in Algorithm~\ref{algo:mmm3}) can be fused into a Node unit stage. The entire dataflow is converted into Fig.~\ref{fig:dataflow_fusion}(a) from its previous design (\figref{fig:dataflow}), applying unique parallelism parameters for each task. Note that the fusion of compute units into a single loop is also data independent. 

One drawback is that the initiation interval might slightly increase, 
due to the merging of multiple stages to form a large stage. Nevertheless, this is worthwhile as fewer stages and boundaries after fusion result in reduced latency and high hardware efficiency.  
\begin{algorithm}[t]
  \caption{Code transformation with an FSM-based structure to transfer an imperfect loop into a perfect loop.}
  \label{algo:loop_fusion}
  \LinesNumbered
  
\begin{footnotesize}

\For(\tcp*[f]{Old}){$i = 0$ to $N_o$  }{ 
\textcolor{blue}{ \st{ \#pragma HLS PIPELINE } }\\
    \For{$k = 0$ to $N_o-1$   } { 
        \textcolor{blue}{ \#pragma HLS PIPELINE} \\
        \textcolor{blue}{ code\_body\_A;}\\
    }
    \textcolor{blue}{code\_body\_B;}\\
}

\vspace{0.2cm}
\For(\tcp*[f]{New}){$i = 0$ to $N_o\times II_{T}$}{ 

    \textcolor{blue}{ \#pragma HLS PIPELINE} \\
    // The loop \textit{II} will be 1 but the equivalent \textit{II} is the \textit{ $II_{T}$ }  \\
    
    \Switch{(curr\_state)}{
        \Case{0}{
            \For{$k = 0$ to Ceiling($\frac{N_o-1}{II_{T}}$) } { 
                \textcolor{blue}{code\_body\_A;}\\
            }
            curr\_state ++;\\
            break;
        }
        \Case{1}{......}
        ......\\
        \Case{($II_{T}-1$)}{
            \For{$k =  (Ceiling(\frac{N_o-1}{II_{T}}) \times (II_{T}-1))$ to $(N_o-1)$ } { 
                \textcolor{blue}{code\_body\_A;}\\
            }
            \textcolor{blue}{code\_body\_B;} \ \ \ \ \ // Used only in this case statement  \\ 
            curr\_state = 0;\\
            break;
        }
       
        \textbf{Default:} break;
    }
}
\end{footnotesize}

\end{algorithm}

\subsection{Handling Imperfect Loops}~\label{sec:fusion_more}
Fusion can potentially result in an imperfect loop, especially in the case of GNNs in which some units operate on edges using an edge-based loop bound while others operate on nodes with a node-based bound. Fig.~\ref{fig:dataflow_fusion}(b) shows the final dataflow once the Edge unit and Node unit are fused into an Edge-Node unit. Algorithm~\ref{algo:loop_fusion} shows a typical example loop from lines 1 to 8 within an Edge-Node unit for GNNs of JEDI-net. The code\_body\_A (e.g., the Concat1 and DNN1 units) iterates with a bound of $N_e$, which is equal to $N_o\times(N_o - 1)$, while the code\_body\_B (e.g., the part of MMM3 from lines 11 to 13 in Algorithm~\ref{algo:mmm3} plus DNN2) only iterates $N_o$ times. After the fusion, it becomes an imperfect loop where the code\_body\_B exists outside the inner loop. To pipeline code\_body\_B, one could set the \#pragma of "PIPELINE" at line 2 in the outer loop, but it leads to automatically unrolling all loops in the hierarchy below. 
As a result, this would completely unroll the inner loop at line 3, resulting in $(N_o - 1)$ hardware copies of code\_body\_A, significantly consuming hardware resources. If the required hardware resources exceed the given budget, one must limit the number of instances of the code\_body\_A. For example, if the total budget can support only $\frac{N_o}{2}$ instances of code\_body\_A, this loop can only be unrolled by a factor of $\frac{N_o}{2}$. 
However, using a \#pragma of unrolling with a factor for the inner loop will not manage to reduce the number of copies since the "PIPELINE" at line 2 has a priority and will force the full unrolling of the inner loop. Thus, one has to move the \#pragma of "PIPELINE" to line 4 to only pipeline the inner loop, excluding pipelining the code\_body\_B, which leads to a poor design and large latency. 

To solve this issue, this work proposes another code transformation which transforms an imperfect loop into a perfect one using a finite-state machine (FSM) based structure with a target initiation interval ($II_T$), as shown from line 9 to 34 in Algorithm~\ref{algo:loop_fusion}. 
If the number of instances of the code\_body\_A that can be deployed under the given resource budget is $N_{A}$, then $II_T$ equals $Ceiling (\frac{N_o-1}{N_{A}} )$. Please note that the increase in the initiation interval is not due to the FSM itself, but is a consequence of limited hardware resources. The loop can now run with an initiation interval of one, but the equivalent initiation interval is the target initiation interval. With this code transformation, we can deploy as many instances as permitted by the hardware budget, thereby improving the design performance after fusion and reducing the overall design latency. An automatic process can be developed to perform the fusion alongside code transformation, effectively handling the imperfect loop and enhancing performance. 


\section{Implementation and co-design framework}~\label{sec:implementation_co_design}
Based on the new low latency hardware architecture incorporating fusion, this section presents a two-level parallelism scheme (Section~\ref{sec:2level_parallelism}) to improve the parallelism and reduce latency. Furthermore, a GNN-specific co-design approach is introduced (Section~\ref{sec:co-design}) to optimize the algorithm and hardware simultaneously. By combining these different optimizations, our designs achieve good performance and efficiency.

\subsection{Two-level Parallelism}~\label{sec:2level_parallelism}
\begin{figure}
\begin{center}
\includegraphics[width=0.9\linewidth]{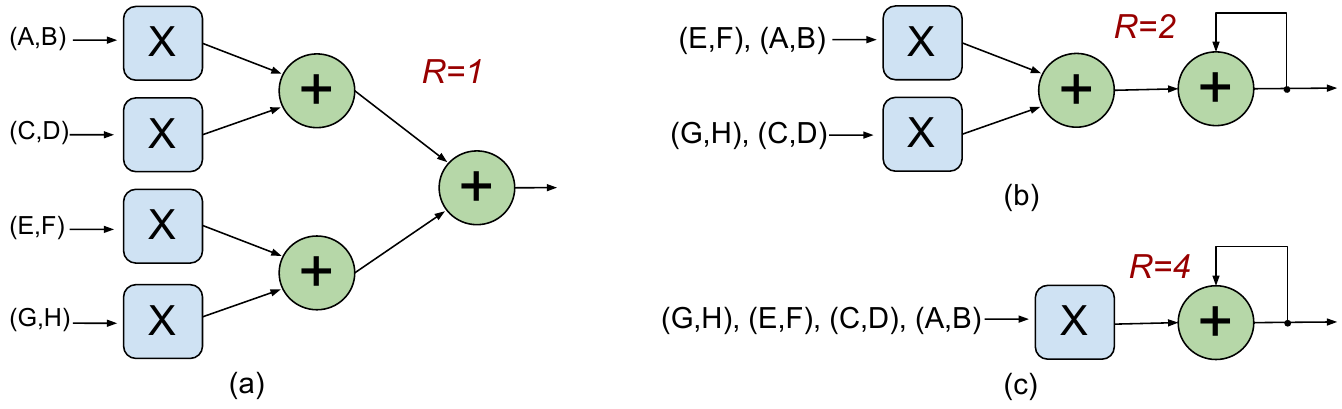}
\end{center}
   \caption{Designs with different reuse factors for a example fully-connected layer which has $N$ multiplications. In this example, the $N$ equals 4. 
   (a) With a reuse factor of 1, all the $N$ multiplications are performed simultaneously using $N$ multipliers, which is fully parallel. 
   (b) With a reuse factor of 2, the 4 multiplications are performed using 2 multipliers. 
   (c) With a reuse factor of 4, the 4 multiplications are performed sequentially using only 1 multiplier, which is fully serial.  
   }
\label{fig:reusefactor}
\vspace{0.2cm}
\end{figure}

The trade-off between latency, throughput and FPGA resource usage is determined by the parallelization of the design. This work exploits a two-level parallelism scheme. First, we utilize the reuse factor~\cite{duarte2018fast} to fine tune the parallelism of computational units within GNNs. The reuse factor controls the amount of parallelism: from sequentially re-using a single multiplier multiple times to using the maximum number of multipliers concurrently. With a reuse factor of one, the computation becomes entirely parallel, and all multiplications are performed simultaneously using a maximal number of multipliers. When the reuse factor is $R$, only $\frac{1}{R}$ of the computation is executed at a time, using $\frac{1}{R}$ fewer multipliers. Fig.~\ref{fig:reusefactor} shows different reuse factors for a fully-connected layer involving four multiplications. The custom code transformation (discussed in Section~\ref{sec:sparsity}) is performed to optimize the matrix multiplications to avoid multiplications. Hence, only the three MLPs ($f_R$, $f_O$, $\phi_O$) require multipliers in the JEDI-net design. We apply the reuse factors $R_{fR}$, $R_{fO}$ and $R_{\phi O}$ to these three MLPs. This work always tries to achieve extremely low latency by using as many hardware resources as feasible, such as unrolling all the layers in the MLPs by adopting a reuse factor value of 1.  

In addition, we deploy multiple instances of the $f_R$ unit to further increase the design parallelism. The $f_R$ is applied to each column of the $B$ matrix, as described in~\secref{subsec:jedi-net}, resulting in a significant number of iterations since there are $N_o \times (N_o-1)$ columns in the $B$ matrix. Completely unrolling all the iterations requires thousands of hardware copies of $f_R$, leading to considerable hardware resource consumption that could easily exceed a given FPGA. Hence, this work partially unrolls it with a factor, $N_{fR}$, producing $N_{fR}$ hardware instances of the $f_R$ unit, each processing a vector of the $B$ matrix. The preceding hardware units of $f_R$ should be also updated so that they can produce $N_{fR}$ vectors in each cycle. Moreover, the succeeding hardware unit, the aggregate unit (MMM3), is also designed to be able to consume multiple result vectors as discussed in Section~\ref{sec:outer_mmm}. 

%

\subsection{Resource Model}~\label{sec:resource_model}
To explore design space and identify the best trade-off between hardware resources and design performance, a resource model for GNN-based JEDI-net has been developed. Since the DSPs are the most critical resources for FPGA-based neural networks, our resource model primarily focuses on DSPs. The DSP resource model is detailed in Equation~\eqref{eq:resource}:

\begin{footnotesize}
\begin{align}
 DSP_{layer} &= \frac{ FC_{in} \times FC_{out} }{R_{NN}}  \nonumber
 \\
 DSP_{NN} &= \sum_{layer = 1}^{total} DSP_{layer} \label{eq:resource} 
 \\
 DSP_{model} &= \sum_{NN={f_R, f_O,\phi_O}} DSP_{NN} \times N_{NN} \leq DSP_{total} \nonumber
\end{align}
\end{footnotesize}
Here, $DSP_{layer}$ denotes the number of DSPs utilized for a single fully-connected (FC) layer, with $FC_{in}$ and $FC_{out}$ representing input and output sizes, respectively. 
The JEDI-net has three MLPs, each consisting of multiple FC layers. $NN$ represents the labels of these three MLPs, namely $f_R, f_O$ and $\phi_O$, as shown in~\figref{fig:sys}(b). For simplicity, this work utilizes a uniform bitwidth (Q12.12, 24 total bits with 12 fractional bits) for most of the design datapath. However, as the weight values of the MLPs fall in the range [0,1), approximately 13 effective bits are used. Hence, the Vivado HLS tool trims the remaining bits from one of the inputs for the multipliers, allowing a single Xilinx DSP to fully implement one multiplier.
Furthermore, we use Q16.16 for the accumulators to maintain accuracy. 
With our proposed approach, there are no multipliers in MMM1/2/3 units. As a result, only the MLP units need multipliers which are implemented using DSPs. The total number of DSPs used in JEDI-net, as shown in Equation~\eqref{eq:resource}, should be less than the total number of DSPs available on the targeted FPGA. 

\subsection{Latency Model}\label{sec:laten_model}
We have managed to fuse most of the sub-layers, which leads to a lower latency architecture compared to the one proposed in our previous work~\cite{que2022optimizing}.
The latency model of JEDI-net based on this latency optimized hardware architecture is detailed in Equation~\eqref{eq:II}.

\begin{footnotesize}
\begin{align}
II_{loop} &= II_{mult} \times max(Ceiling(\frac{N_O-1}{N_{fR}}), R_{fO}, R_{\phi O} )  \nonumber
 \\
 II_{model} &= II_{loop} \times N_O  \label{eq:II} 
 \\
 Latn_{model} &= II_{loop} \times (N_O - 1) + DP_{loop} + DP_{tail} \nonumber
\end{align}
\end{footnotesize}
Here, $II_{mult}$ represents the initiation interval of the multiplier, which is one cycle in this work. $II_{loop}$ denotes the initiation interval of the fused loop, which depends on the maximum number of $f_R$ units that can be deployed on a given FPGA with limited hardware resources, and reuse factors for $f_{fO}$ and $f_{\phi O}$. Please note that $R_{fR}$ is always set to 1, as the execution of $f_R$ is the bottleneck in the design. Meanwhile, $DP_{loop}$ refers to the pipeline depth of the model while $DP_{tail}$ is the depth of the logic outside the primary fused loop. For simplicity, both of them are constants based on the design architecture. 

\textcolor{\mycolor}{
Equation~\eqref{eq:II} for the proposed hardware architecture is based on~Fig.\ref{fig:dataflow_fusion}(b), incorporating all optimizations introduced in this work. Due to limited hardware resources, the top loop is not unrolled as shown in line 10 of Algorithm~\ref{algo:loop_fusion}. Only
the \#pragma of "PIPELINE" is used. As a result, the Edge-Node unit contains a single node-processing engine, capable of handling one node at a time. The engine, corresponding to the loop body, is designed to be pipelined, allowing for the processing of a subsequent node every $II_{loop}$ (initiation interval) cycles. 
The proposed LL-GNN architecture aims to minimize the loop initiation interval, $II_{loop}$, ideally bringing it down to 1 within the hardware resource budget using the proposed techniques. With increased hardware resources, e.g. a larger FPGA, our architecture can accommodate more engines by fully or partially unrolling the top loop in the Edge-Node unit, increasing the parallelism and further reducing the latency. Furthermore, With a fully unrolled top loop, the latency becomes independent of the number of nodes, given sufficient hardware resources.}

\subsection{Co-design of Algorithm and Hardware of GNNs for Low Latency}~\label{sec:co-design}

\begin{figure}
\begin{center}
\includegraphics[width=0.9\linewidth]{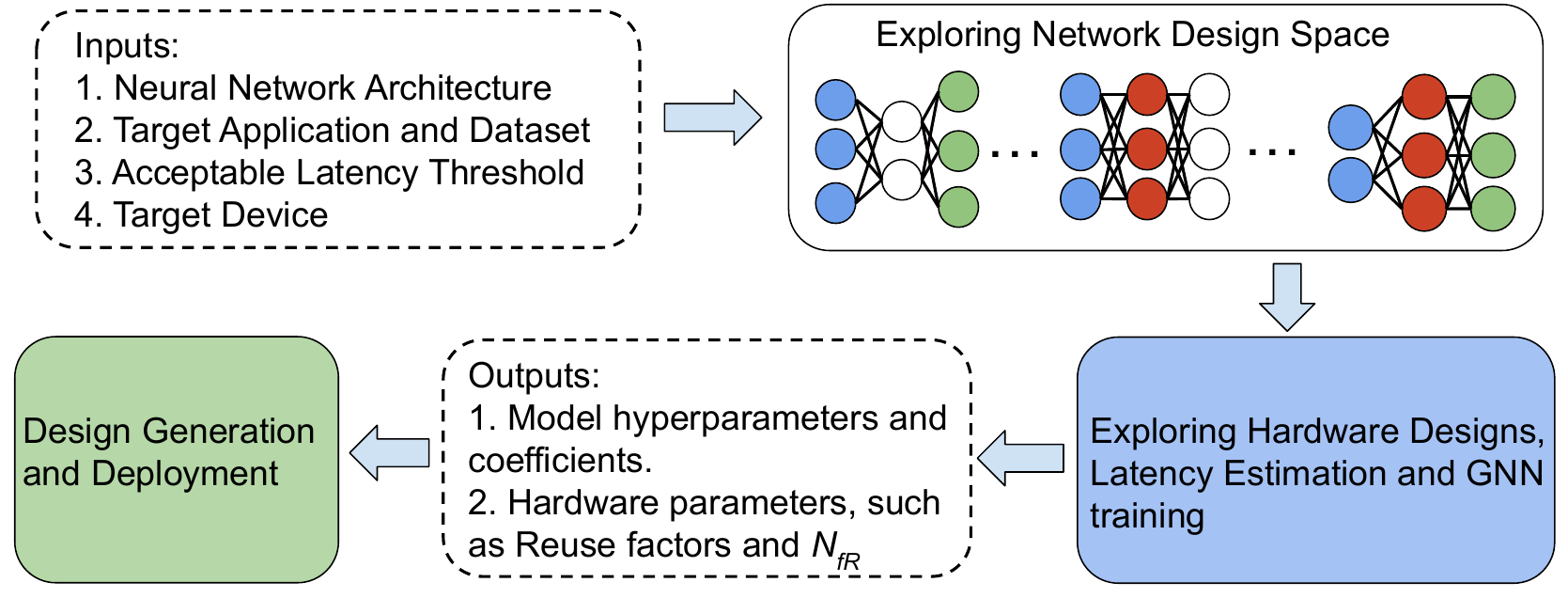}
\end{center}
   \caption{Overview of the co-design approach}
\label{fig:co-design}
\end{figure}

The study in~\cite{moreno2020jedi} focuses on optimizing algorithm for JEDI-net GNNs. However, such strategies often encounter deployment issues where neural network models, optimized for high accuracy, cannot be deployed on a given FPGA due to the large model size or the limited hardware resources. As another line of research, our previous studies~\cite{que2022aicas, que2022optimizing} focus on hardware optimizations. However, without algorithm optimization, it is difficult for these designs to achieve an end-to-end latency of less than $1\mu$s. Thus, existing GNN designs for particle identification either focus on the algorithm or the hardware, leading to sub-optimal designs when deploying GNNs on FPGAs. 

To address this issue, our work proposes a co-design approach tailored for GNNs, allowing simultaneous optimization of both the algorithm and hardware. This strategy enables us to explore trade-offs within user-defined constraints, such as maximizing accuracy with a latency requirement of less than 1$\mu$s on a given FPGA. An overview of our proposed co-design approach is shown in Fig.~\ref{fig:co-design}. The co-design approach consists of the following key steps:
\begin{itemize}

\item \textbf{Define Acceptable Latency Threshold}: 
Given that this work focuses on achieving low latency on a resource-constrained hardware device with a strict latency requirement ($Latn_R$), we further refine our design search space and exploration flow. We introduce a unique parameter, $\alpha$, to fine-tune an acceptable latency threshold. Models whose latencies ($Latn$) exceed this threshold (defined as $\alpha \times Latn_R$) are excluded from the training process. This early termination strategy avoids the unnecessary training of models that would be unfit for deployment due to latency constraints. By carefully choosing $\alpha$, we provide a balance between training time and exploration of the solution space. 

\item \textbf{Rebalance MLP Sizes}: Recognizing that edge-related operations usually require more iterations than those involving nodes due to the larger number of edges in a graph, we rebalance the sizes of different MLPs when exploring network design space. For example, we assign a smaller value to the hidden layer size ($S_{fR}$) in the edge-related function $f_R$, while maintaining or increasing the size ($S_{fO}$) of the other two MLPs. It is because the $f_R$ is required to run $N_O\times(N_O-1)$ times, which is significantly larger than the $f_O$ that runs only $N_O$ times, and the $\phi_O$ that runs only once per inference. This re-balancing reduces latency while maintaining model accuracy.

\item \textbf{Estimate Latency of Design Candidates}: Using equations~\eqref{eq:resource} and~\eqref{eq:II}, we estimate the latency of design candidates with various GNN configurations. The $\alpha$ parameter could be set larger than 1 to loosen the primary constraint and avoid missing potential solutions. $\alpha$ could also be set to a very large value to conduct a full coverage.

\item \textbf{Generate Optimal Design}: Once we find the optimal design based on the user-defined metrics, the model hyperparameters, as well as weights and bias, are generated. A low-latency FPGA design is then produced using our open-source HLS-based templates, thereby enhancing design productivity. 

\end{itemize}

Although~\cite{moreno2020jedi} has conducted a wide algorithmic search considering various model hyperparameters, it assigns uniform size to all three MLPs, which is not latency friendly for GNNs since edge-related operations usually require more iterations than node-related ones as discussed above. We mitigate this by rebalancing the sizes of different MLPs, thereby striking a balance between algorithm and hardware performance. \textcolor{\mycolor}{
In design space exploration (DSE), we also use grid search, like~\cite{moreno2020jedi}, to navigate both algorithmic and hardware design parameters, constrained by latency considerations. We recognize that more advanced DSE techniques, such as those leveraging reinforcement learning, might offer better efficiency over grid search. We leave this for our future work since it has limited impact on the conclusions in this paper.
}

While our primary objective in this work is to minimize latency on a resource-constrained device, the optimization mode of our co-design approach can easily be adapted  to serve other user-defined metrics and constraints. Furthermore, we are investigating how this co-design approach can be automated by techniques such as meta-programming~\cite{vandebon2021enhancing, que2023metaml}.

\subsection{System Overview}

\begin{figure}
\begin{center}
\includegraphics[width=1.0\linewidth]{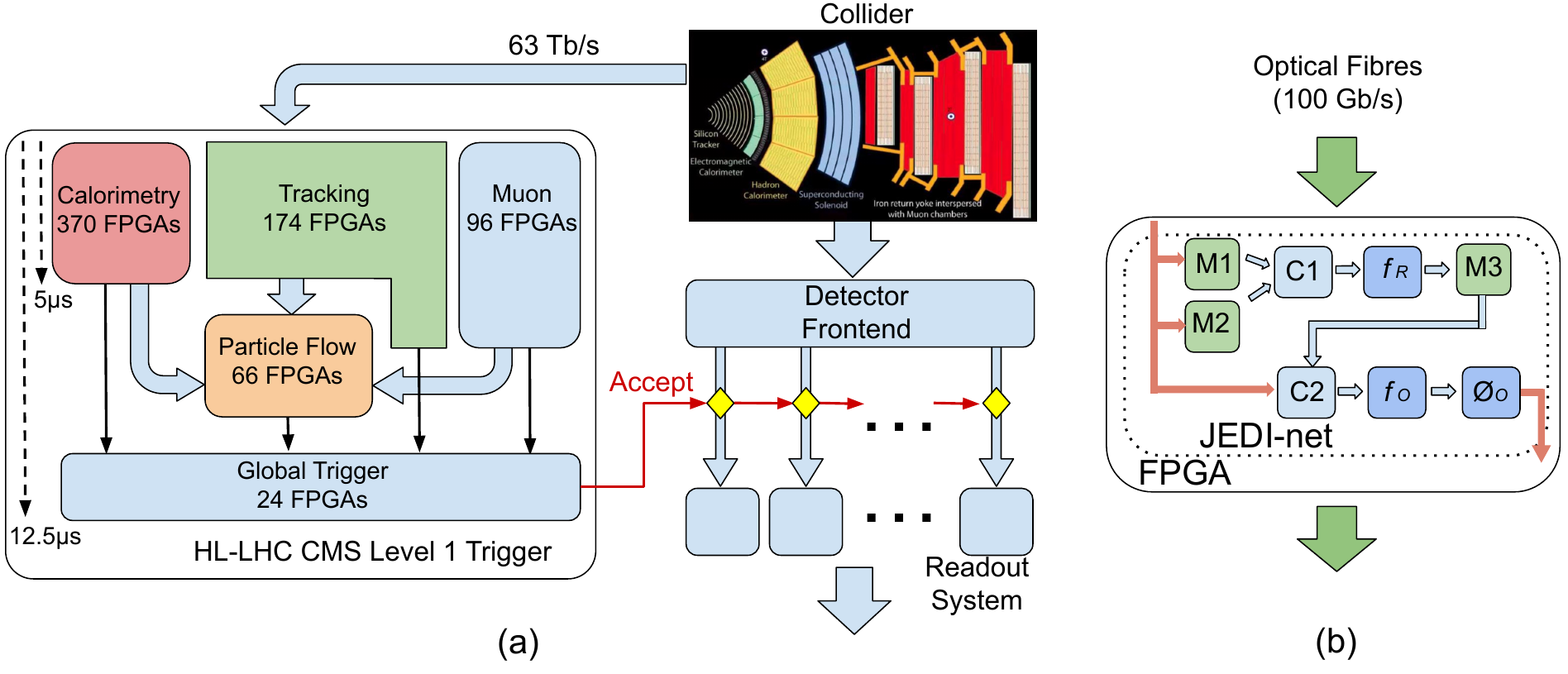}
\end{center}
   \caption{\textcolor{\mycolor}{System Overview. (a) The CMS L1T as well as data acquisition system;} (b) The FPGA system of JEDI-net for particle detection. }
\label{fig:fpga}
\end{figure}

The Compact Muon Solenoid (CMS) serves as one of the particle detectors at CERN LHC. The CMS Experiment, which is related to the triggering and collisions at this particle detector, utilizes two levels of real-time triggering. 
The data acquisition system, including the L1T, is shown in \figref{fig:fpga}~(a). 
The L1T is responsible for making fast, real-time decisions about which events from the LHC should be kept for further examination. 
It works at an input rate of 40 MHz, matching the rate of proton-proton collisions at the LHC, resulting in a data rate of hundreds of terabytes per second. Given the limitations in readout bandwidth and storage capacity, it is not feasible to store all the data produced. 

The aim of the L1T, therefore, is to decrease the data rate by an average of 400 times, 
from 40 MHz to approximately 100 KHz, 
a rate that can be efficiently transferred to the data storage for later analysis in detail.
\textcolor{\mycolor}{In order to provide the required scientific performance, the HL-LHC L1T system relies on the delivery of highest precision inputs representing the largest bandwidth ever handled at Level-1 (63 Terabits per second of information)~\cite{CERN2020L1T}.
The significant enhancement of the L1T in the upgraded HL-LHC compared to the previous LHC is the implementation of the Particle Flow unit that aggregates the regionally processed information from various sub-detectors, including Calorimetry, Muon and Tracking, and facilitates sophisticated algorithms, such as GNN-based JEDI-net, to make a more accurate decision for L1T.
If an accept decision is made, the data are then read out for further processing; otherwise, the data will be dropped. }

The L1T has a total latency budget of 12.5$\mu$s in the HL-LHC~\cite{CERN2020L1T}. \textcolor{\mycolor}{The 12.5$\mu$s is a fixed latency (there is no batching), with a hard limit (we lose the data if exceeded). It consists of numerous FPGA-based subsystems with around 730 FPGAs.} They are designed without a CPU or PCIe in the data path to meet ultra-low latency requirements. Furthermore, no external memory is included in the data path.
\textcolor{\mycolor}{One of the candidate algorithms for the HL-LHC is the GNN-based particle identification algorithm. For a single algorithm the constraint of latency is 1$\mu$s. The GNN designs accept AXI bus based streamed data which arrive via multiple parallel optical fibres, each running at 25 Gbps, as shown in~\figref{fig:fpga}~(b).} 
\textcolor{\mycolor}{Dedicated transceiver units facilitate the reception and transmission of data packets to and from the FPGA, ensuring that data flow is maintained at the speeds required for real-time processing in HL-LHC. Besides, the data links use a custom protocol with no forward error checking, hence they are able to run with a latency of $\sim$100 ns.
Since the data links run with such low latency, there are no concerns about the latency of incoming data. Moreover, this link latency can often be effectively overlapped with the data processing, for example, in this work the initiation intervals of the proposed designs are between 150ns to 750ns. Therefore, the link latency is excluded in the evaluations of algorithms. This is compatible with the latency calculation for algorithms and applications in CERN HL-LHC experiments and aligning with other designs~\cite{coelho2021automatic, duarte2018fast} of particle identification for the L1 trigger system.}

\section{Evaluation and Analysis}
This section presents the evaluation results of the GNN-based JEDI-net on FPGAs demonstrating the scalability of the proposed optimization for GNNs. 

\subsection{Experimental Setup}
This study covers JEDI-net-30p~\cite{moreno2020jedi} and JEDI-net-50p GNN models, targeting datasets of 30 particles~\cite{jet_dataset_30p} and 50 particles~\cite{jet_dataset_50p}, respectively. 
To study the performance and limitations of the proposed optimizations and hardware architecture, the designs are implemented using Xilinx Vivado HLS \textcolor{\mycolor}{2019.2} on a Xilinx Alveo U250 board with an UltraScale+ XCU250 FPGA for the evaluation and comparison with other implementations. It offers 1.7M LUTs, 12.3k DSP slices and two 100Gbps network interfaces. 
It runs at 200MHz so each cycle is 5ns. 
The hardware datapath employs a Q12.12 fixed-point representation with 1 sign bit, 11 integer bits and 12 fractional bits. In contrast, the accumulator uses Q16.16 to keep accuracy.
It achieves the same accuracy as the floating-point model.




\subsection{Balancing Quantization and Accuracy}

\begin{figure} 
    \subfloat[]{%
    \includegraphics[width=0.52\linewidth]{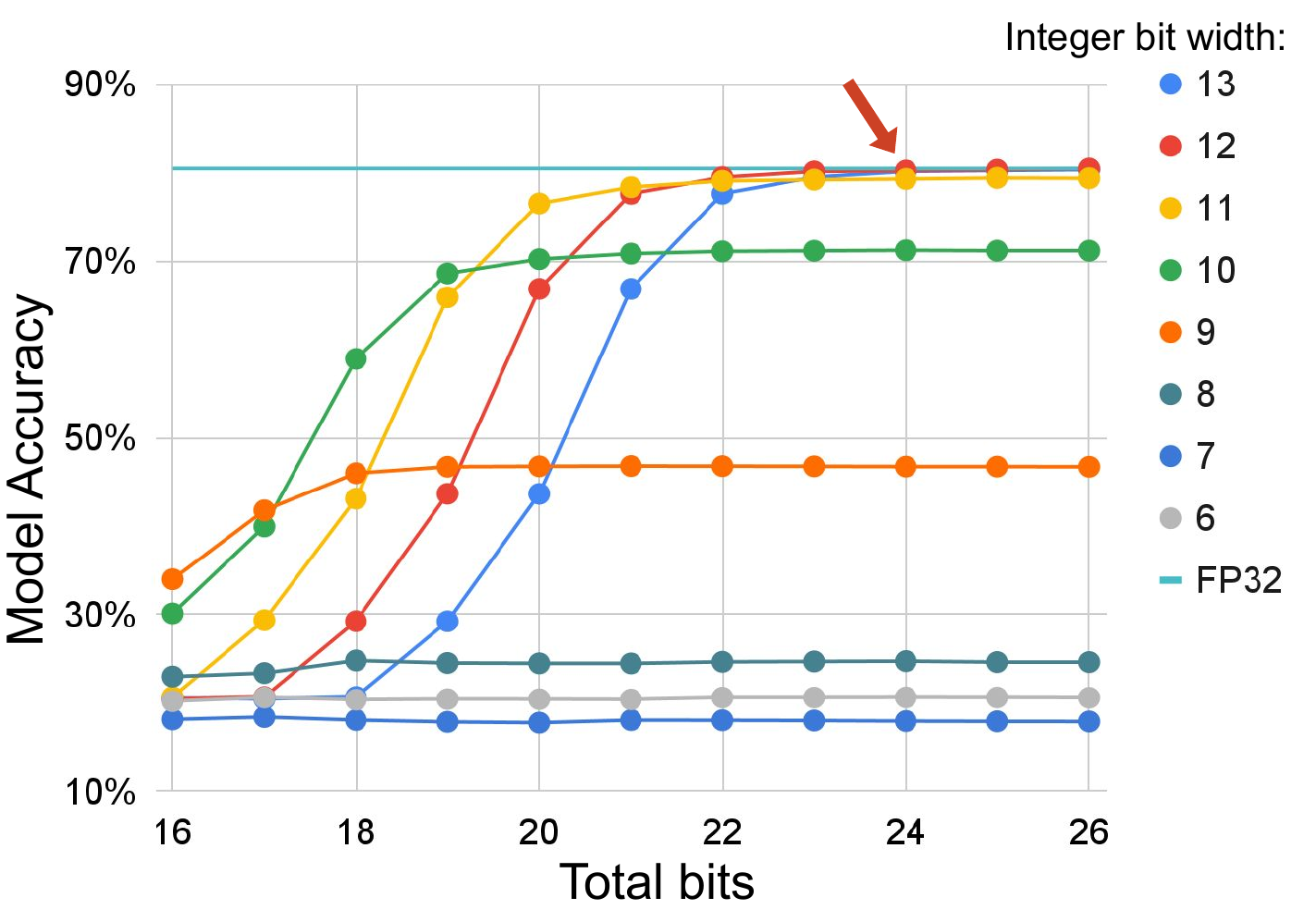}}
  \hfill
    \subfloat[]{%
    \includegraphics[width=0.48\linewidth]{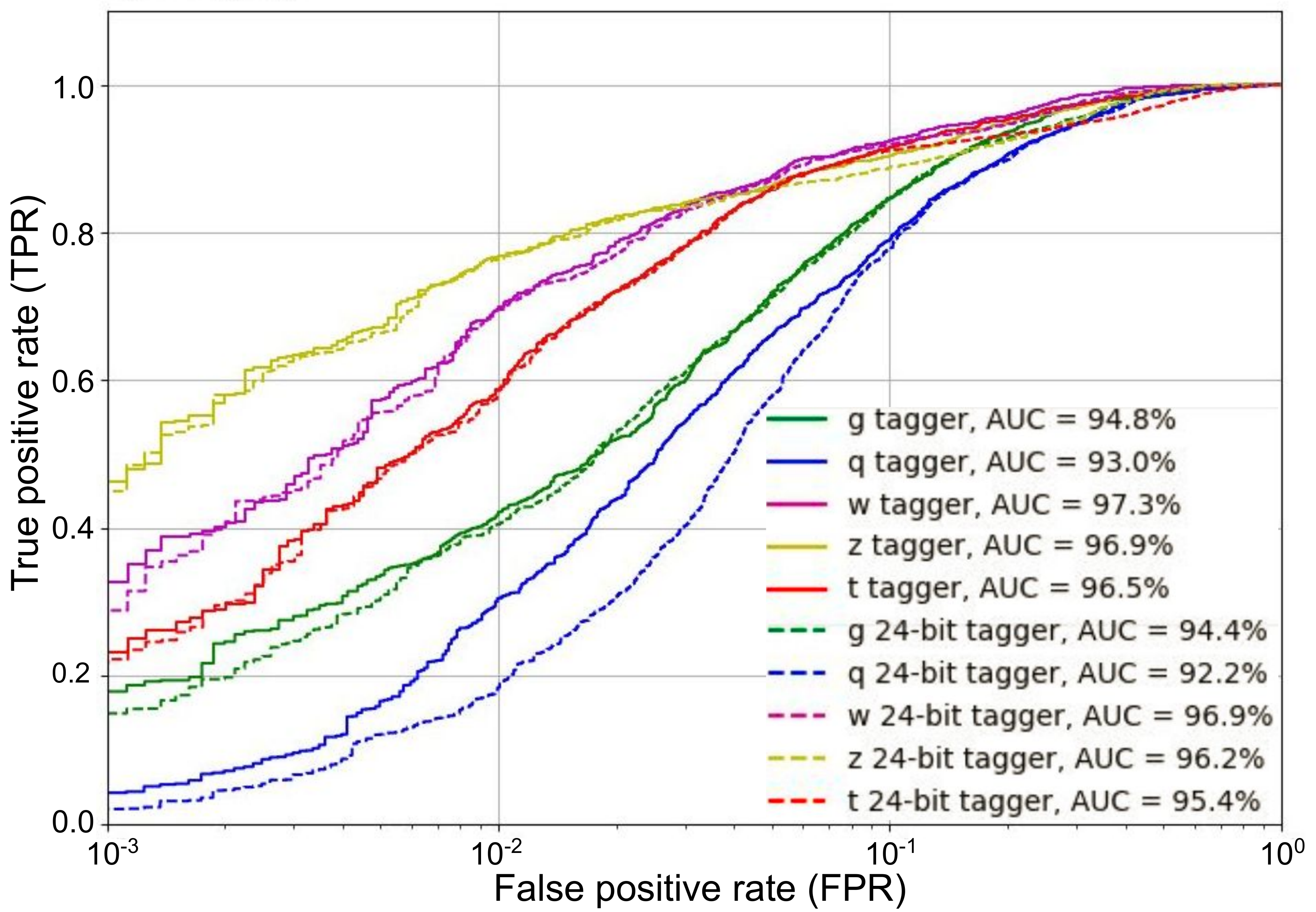}}
  \vspace{0.2cm}
  \caption{(a) Model accuracy of JEDI-net 50p network under different datapath bitwidth configurations. The x-axis represents the total number of bits, while the different colored lines with dots represent varying numbers of integer bits. The line without dots shows the model accuracy using 32-bit floating-point (FP32) representation. 
  (b) The ROC curves with the AUC for the 5 jets, including gluon, light quarks, W boson, Z boson and top quark. }
  \vspace{0.2cm}
  \label{fig:accuracy_auc} 
\end{figure}

To find an optimal fixed-point representation that can achieve no reduction in the physics performance of the algorithm compared to the conventional floating-point representation, we analyze the fixed-point precision across a range from 16 to 26 bits in total bit widths and 6 to 13 integer bits (including the sign bit), as shown in Fig.~\ref{fig:accuracy_auc}(a). 
With 24 bits in total and 12 integer bits, the fixed-point model effectively achieves the same accuracy as the FP32 floating-point counterpart. In addition, JEDI-net achieves much higher accuracy than the previous work based on MLPs~\cite{duarte2018fast,coelho2021automatic} which only has an accuracy of around 75\%.

To further understand the performance of our optimized model, we also evaluate the Receiver Operating Characteristic (ROC) curves with the area under the curve (AUC) for the 5 jet classifiers, including gluon, light quarks, W boson, Z boson, and top quark. These results are shown in Fig.~\ref{fig:accuracy_auc}(b), with higher AUC value corresponding to superior classification performance. Although the AUC of the light quarks tagger (blue lines) using 24-bit fixed-point data representation seems different from the floating-point one, it's important to note the logarithmic scale on the x-axis of Fig.~\ref{fig:accuracy_auc}(b) and the AUC loss of the q tagger is less than 1\%, thus demonstrating that the fixed-point representation has a negligible impact on the overall performance. In summary, we successfully balance the trade-off between quantization and accuracy, thereby maximizing the efficiency and performance of our optimized model.





\subsection{Resource Utilization}

\begin{table}
\centering
\caption{Resource utilization.}
\label{table:utili}

\scalebox{0.9}{\begin{tabular}{ c | c | c |c| c| c}
\toprule
 \multirow{2}{*}{\textbf{Task}} & & \textbf{LUT}   & \textbf{FF}  & \textbf{BRAM} & \textbf{DSP}\\
\cmidrule{2-6}
 &  \textbf{Available} & 1728k & 3456k & 5376 & 12288  \\
\midrule
\multirow{2}{*}{JEDI-net 30P (\textbf{J2~\cite{que2022optimizing}})}
& Used [$\downarrow$] &  1158k &  246k & 1392 & 11504 \\ \cmidrule{2-6}
& Utilization [\%, $\downarrow$] & 67 &  7.1 & 25 & 93  \\
\midrule
\multirow{2}{*}{JEDI-net 30P (\textbf{J3}, w/ fusion)}  
& Used [$\downarrow$] &  734k &  137k & 16 & 9013 \\ \cmidrule{2-6}
& Utilization [\%, $\downarrow$] & 42 &  4.0 & 0.3 & 73  \\
\midrule
\midrule
\multirow{2}{*}{JEDI-net 30P (\textbf{J4}, Opt-Latn) }   
& Used [$\downarrow$] &  865k &  138k & 37 & 8776 \\ \cmidrule{2-6}
& Utilization [\%, $\downarrow$] & 50 &  4.0 & 0.7 & 71  \\
\midrule
\multirow{2}{*}{JEDI-net 30P (\textbf{J5}, Opt-Acc) }  
& Used [$\downarrow$] &  911k &   153k & 37 & 9833 \\ \cmidrule{2-6}
& Utilization [\%, $\downarrow$] & 53 &  4.4 & 0.7 & 80  \\
\midrule
\multirow{2}{*}{JEDI-net 50P (\textbf{U4}, Opt-Latn)  }
& Used [$\downarrow$] &  855k &  201k & 25 & 8945 \\ \cmidrule{2-6}
& Utilization [\%, $\downarrow$] & 49 &  5.8 & 0.5 & 73  \\
\midrule
\multirow{2}{*}{JEDI-net 50P (\textbf{U5}, Opt-Acc)   }
& Used [$\downarrow$] &  815k &   189k & 37 & 8986 \\ \cmidrule{2-6}
& Utilization [\%, $\downarrow$] & 47 &  5.5 & 0.7 & 73  \\
\bottomrule
\end{tabular}}
\vspace{0.2cm}
\end{table}

Table~\ref{table:utili} shows the resource utilization of our designs on the U250 FPGA with different configurations. Arrows in the table symbolize desired trends. 
In JEDI-net-30p models, the number of input particle $N_O$ is set at 30. Each particle has a feature size $P$ of 16 as defined by the dataset. For the JEDI-net-50p model, $N_O$ is increased to 50 with the same feature size of $P$. As $N_O$ increases, the number of edges $N_E$ also increases significantly, which is equal to $N_O*(N_O-1)$. This results in 870 edges for $N_O=30$ and dramatically increases to 2450 for $N_O=50$. Both models have various sizes of the hidden layers within $f_R, f_O$ and $\phi_O$ MLPs.   
Table~\ref{table:utili} shows that our newly introduced design with fusion is more resource-efficient than the previous design for the same JEDI-net 30p model. This can be seen clearly when comparing the designs labeled as \textbf{J2} and \textbf{J3} in Table~\ref{table:utili}, both of which target the same JEDI-net model. Further details about \textbf{J2} and \textbf{J3} are provided in~\tabref{table:cmp_u250}. Our new design, based on the proposed low latency hardware architecture with fusion, has reduced the usage of DSPs, BRAM and LUTs. This not only results in lower resource consumption but also paves the way for a higher performance design.

\subsection{Performance Analysis and Discussion}
To show the advantages of our proposed optimizations, \tabref{table:cmp_u250} presents a comparative analysis of various intermediate and final designs derived after applying the proposed optimizations one by one. 
The designs~\textbf{J1$\sim$J5} and \textbf{U1$\sim$U5} correspond to the JEDI-net-30p and JEDI-net-50p networks, respectively, with \textbf{J1} and \textbf{U1} serving as the baselines. 

\subsubsection{\textbf{Creating Baseline Designs}}

Designs \textbf{J1} and \textbf{U1}, targeting the JEDI-net-30p and JEDI-net-50p respectively, have been developed leveraging our proposed outer-product based MMM, custom code transformation of MMMs, and column-major order data representation. These designs are illustrated in Table~\ref{table:cmp_u250} and serve as our baseline designs.
To minimize latency and improve throughput, we further fully unroll the loops within each network layer in the $f_R$, $f_O$, and $\phi_O$ units of designs \textbf{J1} and \textbf{U1}. 
However, the design latency still ranges between 12.56$\mu$s and 32.60$\mu$s, which significantly exceeds the latency constraint of less than 1$\mu$s. Therefore, further optimizations must be explored to meet this hard constraint.

\subsubsection{\textbf{Parallelization and Resource Balancing}}
To further reduce latency and initiation interval, we increase the design parallelism by deploying multiple DNN1 ($f_R$) hardware units. This is particularly necessary since the $f_R$, which serves as the GNN edge function that is applied to each edge in the graph, needs to iterate $N_O*(N_O-1)$ times. 
As a result, as the number of deployed DNN1 units, $N_{fR}$, increases from 1 to 13, the latency of design~\textbf{J2} drops from 12.56$\mu$s (\textbf{J2}) to 1.91$\mu$s, resulting in a 6.5 times reduction in latency. The latency reduction is achieved by allocating more hardware resources to improve performance (low latency). However naively increasing the number of DNN1 units does not guarantee an optimal design. 
It is illustrated in designs~\textbf{U1} and~\textbf{U2} for the large JEDI-net-50p model. Here, latency decreases from 32.60$\mu$s to 12.47$\mu$s as $N_{fR}$ increases from 1 to 3, shown in~\tabref{table:cmp_u250}. However, the required number of DSPs has exceeded the total DSPs on this FPGA. To solve this issue, we strategically re-allocate some DSP blocks from DNN2 ($f_O$) and DNN3 ($\phi_O$) to DNN1 ($f_R$) by increasing the reuse factors of DNN2 and DNN3, as well as deploying multiple DNN1 units to balance the overall design initiation interval as described in Section~\ref{sec:tlp}. This strategy leads to a reduced initiation interval and latency. A subsequent design space exploration helps determine the appropriate values of the parallelism parameters $(R_{fO}, N_{fR})$ to be (4,4). As a result, we arrive at design~\textbf{U3} which has both better initiation interval and latency compared to the design~\textbf{U2}. 

\begin{table}[t]
\centering
\caption{Performance comparison of FPGA-based implementations of JEDI-net with various parameters.}
\label{table:cmp_u250}

\scalebox{0.88}{\begin{tabular}{c|c|c|c|c|c||c|c|c|c|c}
\toprule
  & \textbf{J1} &\textbf{J2}  &\textbf{J3}  &\textbf{J4} &\textbf{J5} & \textbf{U1} & \textbf{U2} & \textbf{U3} & \textbf{U4} & \textbf{U5}  \\ 
\midrule
  & \cite{que2022aicas}  & \cite{que2022optimizing}   & \multicolumn{3}{c||}{ This work } & \multicolumn{3}{c|}{ \cite{que2022optimizing} }  & \multicolumn{2}{c}{ This work } \\ 
\midrule
NN Model  & \multicolumn{5}{c||}{ JEDI-net-30p } & \multicolumn{5}{c}{ JEDI-net-50p } \\ 
\midrule
$f_R$ (NL, Size)  & \multicolumn{3}{c|}{ (3, 20) } & (1, 8) & (2, 32) & \multicolumn{3}{c|}{ (3, 50) } & (2, 8)  & (2, 8)\\ 

\midrule
$f_O$ (NL, Size)  & \multicolumn{3}{c|}{ (3, 20) } & (3, 48) &  (3, 48) & \multicolumn{3}{c|}{ (3, 50) }  & (3, 32) & (3, 48)\\ 



\midrule
Accuracy & \multicolumn{3}{c|}{ 78.74\% } & 78.41\% & 79.85\% & \multicolumn{3}{c|}{ 80.42\% } & 80.90\% & 81.18\%\\ 

\midrule
$R_{fO}$  & 1 & 1 & 1 & 1 & 1 & 1 & 1 & 4 & 1  & 1\\ 
\midrule
$N_{fR}$  & 1 & 13 & 10 & 29 & 6 & 1 & 3 & 4  & 25  & 17\\ 
\midrule

\begin{tabular}[c]{@{}c@{}}DSP used \\ \end{tabular} & 
\begin{tabular}[c]{@{}c@{}}1831 \\ (14\%) \end{tabular} & 
\begin{tabular}[c]{@{}c@{}}11504 \\ (93\%) \end{tabular} & 
\begin{tabular}[c]{@{}c@{}}9013 \\ (73\%) \end{tabular} & 
\begin{tabular}[c]{@{}c@{}}8776 \\ (71\%) \end{tabular} & 
\begin{tabular}[c]{@{}c@{}}9833 \\ (80\%) \end{tabular} &
\begin{tabular}[c]{@{}c@{}}7342 \\ (59\%) \end{tabular} & 
\begin{tabular}[c]{@{}c@{}}12894 \\ (104\%) \end{tabular} &
\begin{tabular}[c]{@{}c@{}}12284 \\ (94\%) \end{tabular} &
\begin{tabular}[c]{@{}c@{}}8945 \\ (73\%) \end{tabular} & 
\begin{tabular}[c]{@{}c@{}}8986 \\ (73\%) \end{tabular}
\\

\midrule
II (cycles)  & 880 & 80 & 90 & 30 & 150 & 2462 & 854 & 650 & 100 & 150\\ 
\midrule
II ($\mu$s)  & 4.40 & 0.40 & 0.45 & 0.15 & 0.75 &  12.31 & 4.27  & 3.25 & 0.50 & 0.75 \\ 
\midrule
Latency (cycles) & 2511 & 382 & 124 & 58 & 181 & 6519 & 2493 & 2131  & 130  & 181\\ 
\midrule

Latency ($\mu$s)  & 12.56 & 1.91 & \textbf{0.62} & \textbf{0.29} & \textbf{0.91} & 32.60 & 12.47 & 10.66  & \textbf{0.65} & \textbf{0.91}\\

\midrule
Note  &  &  &  & Opt-Latn & Opt-Acc &   &   &  & Opt-Latn  &  Opt-Acc \\ 
\midrule
\end{tabular}}
\vspace{0.2cm}
\end{table}


\subsubsection{\textbf{Fusion}}

\begin{figure}
\begin{center}
\includegraphics[width=0.7\linewidth]{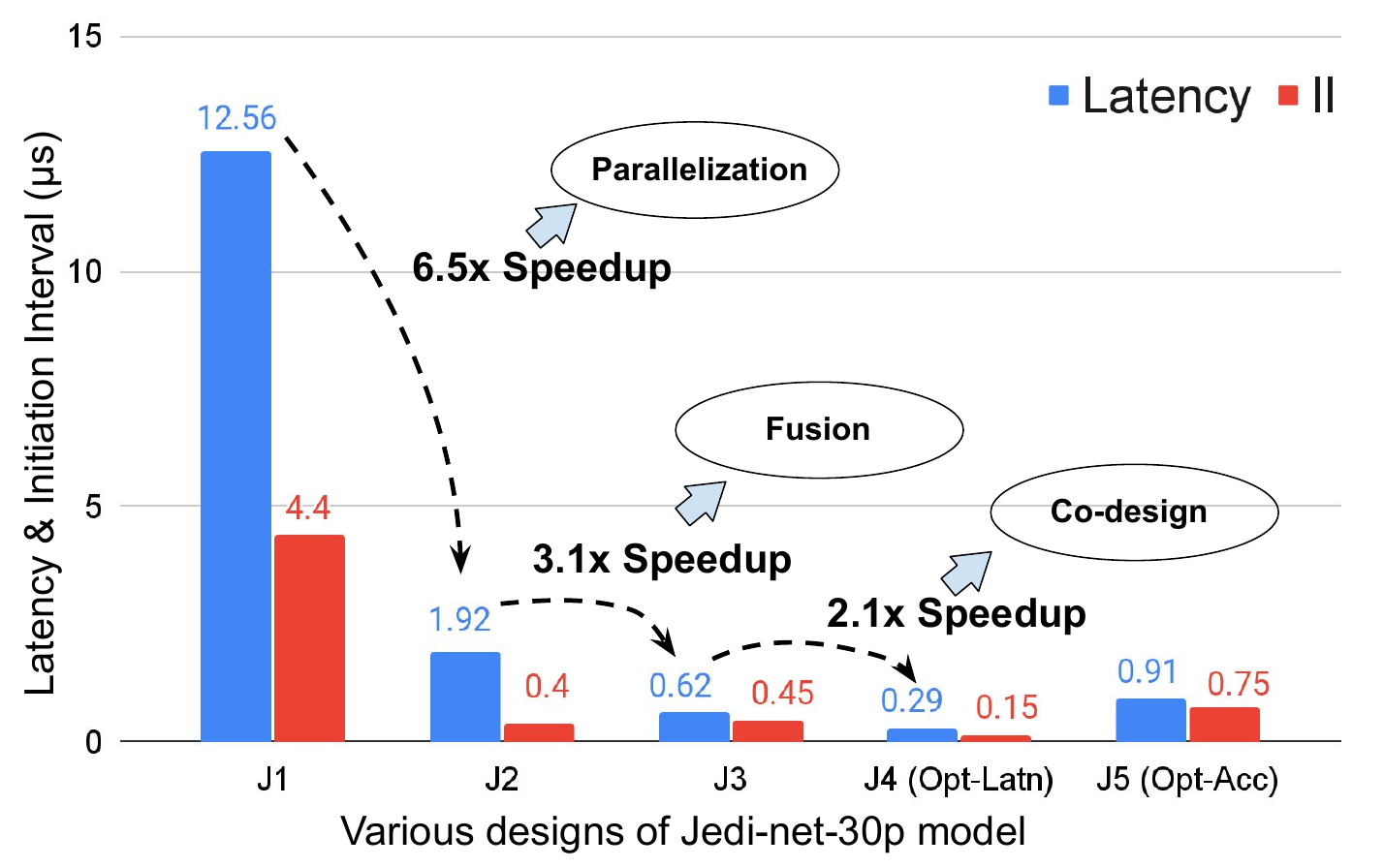}
\end{center}
   \caption{The latency and initiation interval for various designs of JEDI-net-30p models.}
\label{fig:latency_30p}
\end{figure}

The latency can be further reduced by the newly proposed sub-layer fusion. The fusion approach eliminates the boundaries between the hardware units, thus enabling a fine-grained pipeline instead of a coarse grained pipeline. A code transformation with an FSM structure is introduced to mitigate the potential pipeline issue arising from imperfect loops. 
For instance, design \textbf{J3} targets the same neural network model as \textbf{J2}, but with the fusion, the latency is reduced from 1.91$\mu$s to 0.62$\mu$s. This is the first time that the JEDI-net-30p model operates within 1$\mu$s, showing the advantages of the fusion technique in achieving low latency designs. 
Setting $N_{fR}$ to a number between 10 to 14 will lead to the same initiation interval and latency using the new low latency hardware architecture because the partition is performed on the inner loop which has a bound of 29 for JEDI-net-30p. Therefore, we select 10, which demands the least hardware resources. As compared to our previous work \textbf{J2}~\cite{que2022optimizing}, the new design \textbf{J3} is 3.1 times faster as shown in~\figref{fig:latency_30p} while consuming 22\% less DSP blocks. 

These improvements are accompanied by a minor drawback; there is a small increase in the initiation interval from 0.40$\mu$s to 0.45$\mu$s since the pipeline depth of the fused stage increases after fusion. 
However, the small increase in the initiation interval is considered tolerable, especially when considering the significant enhancements in latency and resource efficiency. 

\begin{figure}
\begin{center}
\includegraphics[width=0.7\linewidth]{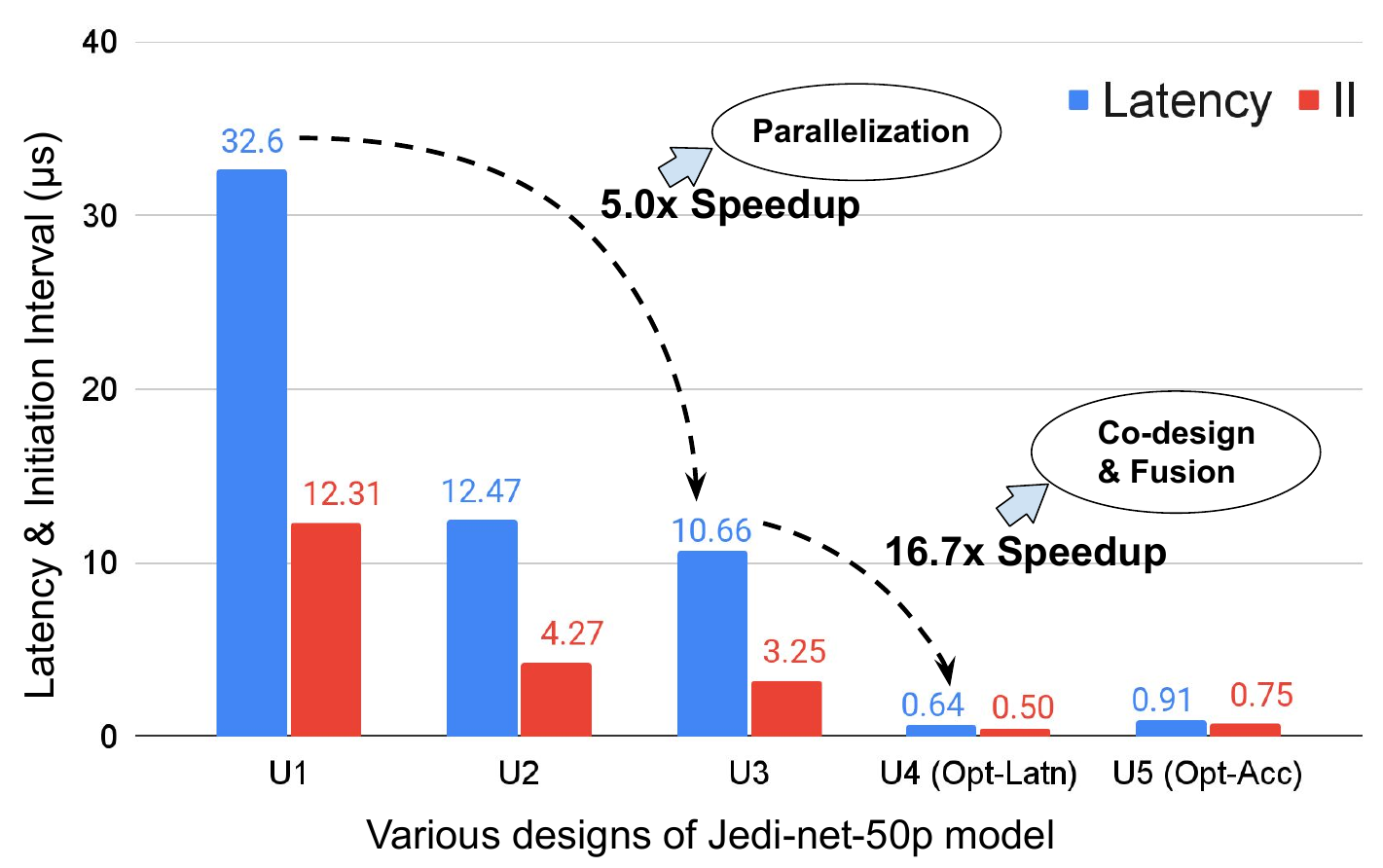}
\end{center}
   \caption{The latency and initiation interval for various designs of JEDI-net-50p models.}
\label{fig:latency_50p}
\vspace{0.2cm}
\end{figure}


\begin{figure} 
    \subfloat[]{%
    \includegraphics[width=0.48\linewidth]{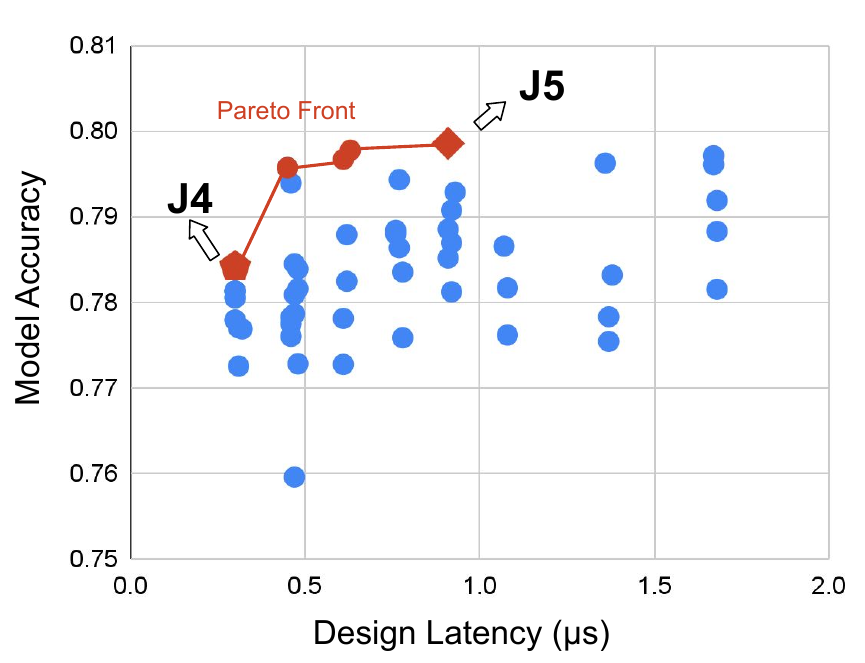}}
  \hfill
    \subfloat[]{%
    \includegraphics[width=0.48\linewidth]{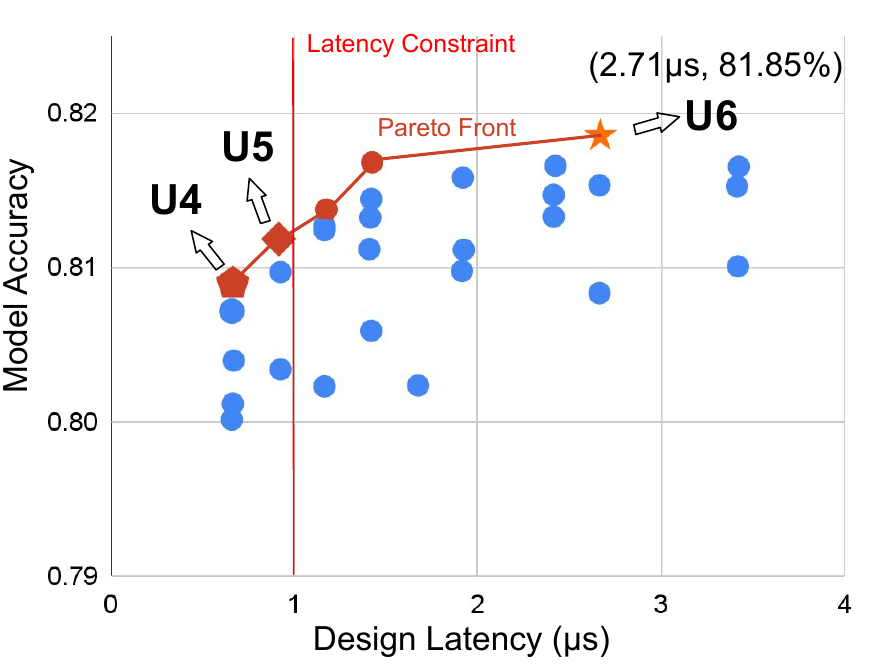}}
  \caption{(a) The latency and accuracy for JEDI-net-30p models.
  (b) The latency and accuracy for JEDI-net-50p models.  }
  \vspace{0.2cm}
  \label{fig:latency_accuracy} 
\end{figure}

\subsubsection{\textbf{Algorithm and Hardware Co-optimization}}
We introduce a co-design approach to optimize the GNNs both on model accuracy and hardware performance. It is worth noting that this is the first instance where the larger JEDI-net-50p models can run in 1$\mu$s without sacrificing model accuracy. 
In~\cite{moreno2020jedi} and our previous work~\cite{que2022aicas, que2022optimizing}, all the MLPs share the same size. 
However, our analysis of JEDI-net reveals that the $f_R$ is iterated $N_O$ times more than $f_O$ and $N_O*(N_O-1)$ times more than $\phi_O$. As a result, if $f_R$ is smaller, more copies of $f_R$ units can be deployed on a given FPGA, leading to a faster design. 
We consider JEDI-net-30p models with the number of layer (NL) of $f_R$ searched in (1, 2, 3, 4) and layer size in (8, 16, 24, 32). The latency parameter $\alpha$ is set to 2 for JEDI-net-30p. As JEDI-net-50p is much larger, the size of $f_R$ is searched in (8, 16, 32, 48) with NL in (1, 2, 3, 4). The $\alpha$ is set to 4 for JEDI-net-50p. For simplicity, we keep the layer number and other configurations of $f_O$ and $\phi_O$ the same as in~\cite{moreno2020jedi} but adjust only the size of their first layer to one of (16, 32, 48, 64, 96) for both JEDI-net-30p and 50p. 

The results are shown in~\figref{fig:latency_accuracy}(a) and (b) using a U250 FPGA platform. Each blue dot in these figures represents an explored design. \figref{fig:latency_accuracy}(a) shows the designs with a latency less than 2$\mu$s. Since JEDI-net-50p is much larger than JEDI-net-30p, \figref{fig:latency_accuracy}(b) shows the designs with a latency less than 4$\mu$s.
Designs \textbf{J4} and \textbf{U4} are selected because they have the lowest latency (Opt-Latn) among all the candidate designs and they have the highest model accuracy among the designs with the same latency. Despite a minor accuracy loss of 0.33\%, the latency of JEDI-net-30p can be further reduced to 0.29$\mu$s, which is 2.1 times faster than~\textbf{J3}. The initiation interval can also be reduced to 0.15$\mu$s. As for JEDI-net-50p, the latency of \textbf{U4} is reduced from 10.66$\mu$s to 0.65$\mu$s, which is 16.7 times faster than \textbf{U3}~\cite{que2022optimizing}, as shown in~\figref{fig:latency_50p}. In addition, \textbf{U4} maintains higher accuracy than \textbf{U3}.


\subsubsection{\textbf{Analysis and discussion}}
Although our focus is on low latency GNNs, for a real-world application we should also consider the model accuracy. We select designs \textbf{J5} and \textbf{U5} as they have the highest model accuracy (Opt-Acc) among all the designs with a latency less than 1$\mu$s. Compared to our previous work~\cite{que2022optimizing} (\textbf{J2}), the new design \textbf{J5} achieves 1.11\% better model accuracy and 2.1 times reduction in latency. For JEDI-net-50p, the newly searched design \textbf{U5} achieves 0.76\% better model accuracy and 11.8 times reduction in latency compared to the previous work (\textbf{U3}). Design \textbf{U6}, another design for JEDI-net-50p, is an interesting case that achieves the highest accuracy among all the searched designs, but its latency is 2.71$\mu$s which exceeds the latency constraint ($1\mu$s) for the CERN HL-LHC. However, with the possibility of using a larger FPGA chip in the future, this latency could be further reduced to an accepted level through our optimizations. 
In addition, based on the latency model outlined in~\secref{sec:laten_model}, the estimated latency of designs \textbf{J4, J5, U4} and \textbf{U5} are 0.30$\mu$s, 0.91$\mu$s, 0.66$\mu$s and 0.915$\mu$s respectively, leading to prediction errors of less than 5\% . 

We demonstrate that our approach can not only find a design with a much better latency (Opt-Latn) but can also find a more balanced design, such as a design achieving better accuracy with a higher but still acceptable latency ($< 1\mu$s), such as~\textbf{U5}. 
Moreover, future detectors will require higher sensitivity and lower latency. 
Our proposed approach and optimizations can be used in developing designs in next-generation FPGA technology to meet such requirements.




When particle identification is part of the data processing in collider trigger systems, our approach can still lead to an appropriate set of parameters to get the optimal initiation interval and latency with a given hardware budget. 
Besides, in a realistic use case, the cardinality of the input dataset, \textcolor{\mycolor}{such as the number of features in an input vector,} might be much smaller. In that case, one would be able to speed up the algorithm even more than what we show in this work, as well as to reduce the resource utilization.

\textcolor{\mycolor}{When the DSP resources on the given FPGA are insufficient, the bind\_op pragma offers designers the option to implement an operation using LUTs instead of DSPs. This could potentially expand the design space and alleviate the bottleneck associated with DSPs. We will leave this for our future work as our LUT utilization is not low, and it doesn't impact the conclusions drawn in this work.}

The proposed approach adopts task-level parallelism and dataflow architecture. It is also suitable for deployment on multiple heterogeneous compute devices. To ensure efficient execution, tasks can be allocated to different compute devices with the intra-task and inter-task optimizations based on the features of the task, the strengths and limitations of the devices as well as the connection structure of the devices.


\subsection{Comparison with GPUs and CPUs}
To compare the performance of the proposed design on FPGA with other platforms, we run the JEDI-net models implemented in~\cite{moreno2020jedi} on Intel Xeon Gold 6154 CPU and NVIDIA GeForce RTX 3090 (CUDA 11.8) based on PyTorch (1.12.1) framework. The CuDNN libraries are used for optimizing the hardware performance on GPUs. Each batch has 1000 graph events (samples) according to~\cite{moreno2020jedi}, so we set the same batch size across all the hardware platforms for a fair comparison.
CPU power consumption is measured by the pcm-power utility~\cite{han2017ese}, excluding the DRAM power consumption. GPU power consumption is measured using nvidia-smi utility. \textcolor{\mycolor}{The maximum total power consumption of the U250 board is used in comparison, which is 225W.}

We adopt KGPS (Kilo Graphs Per Second), which denotes the number of graph events inferences that run per second, as an indicator of throughput. 
Compared with the JEDI-net implementation on GPU, our FPGA design is 8.7$\sim$9.0 times faster.
In terms of the power efficiency given by KGPS per Watt,
our design is \textcolor{\mycolor}{12.6$\sim$13.1} times higher than the GPU implementation.
When compared to the CPU implementation, our FPGA implementation is 282$\sim$545 times faster. In addition, our design achieves \textcolor{\mycolor}{141$\sim$296} times higher power efficiency than the CPU implementation. As for future improvements, we believe the proposed custom code transformation can also be applied to CPU and GPU implementations, but the latency profiling shows that the three MMMs cost less than 15\% of the total latency. We leave that for future work since it has a limited impact on the conclusions in this paper. 
The FPGA implementation is faster and more efficient due to tailor-made optimizations and a co-design approach. 
\begin{table}[t]
\centering

\caption{\textcolor{\mycolor}{Comparison of the FPGA, CPU and GPU designs.}}
\label{tab:cmp_cpugpu}
\begin{threeparttable}
\centering
\scalebox{0.95}{
\begin{tabular}{c|c|c|c|c|c|c}
\toprule

Platform
& \multicolumn{2}{c|}{ CPU Gold 6154}
& \multicolumn{2}{c|}{GPU GeForce 3090}
& \multicolumn{2}{c}{FPGA U250 }
\\ \midrule

Frequency 
&\multicolumn{2}{c|}{3.00 GHz}
&\multicolumn{2}{c|}{1.70 GHz}
&\multicolumn{2}{c}{200 MHz}
\\
\midrule
Technology 
&\multicolumn{2}{c|}{14 nm}
&\multicolumn{2}{c|}{8 nm}
&\multicolumn{2}{c}{16 nm}
\\ \midrule
Precision 
&\multicolumn{2}{c|}{F32} 
&\multicolumn{2}{c|}{F32} 
&\multicolumn{2}{c}{24 Fixed}   
\\ 
\midrule

GNN Model
& 50p\tnote{1} & 30p\tnote{1} 
& 50p\tnote{1} & 30p\tnote{1} 
& 50p ( U4 )
& 30p ( J4 )
\\

\midrule
Accuracy (\%) 
& 80.90 
& 78.41 
& 80.90 
& 78.41
& 80.90 
& 78.41
\\ 

\midrule
Power (W) 
& 106
& 112  
& 325
& 326
& \textcolor{\mycolor}{225} 
& \textcolor{\mycolor}{225} 
\\ 
\midrule
Batch Size & \multicolumn{6}{c}{1000} \\ 
\midrule
Average Lat. ($\mu$s)
& 272.4
& 42.3  
& 4.5
& 1.3 
& 0.50 
& 0.15 \\ 
\midrule
Throughput (KGPS) 
 & 3.7   
 & 23.6 
 & 222.2
 & 769.2  
 & 2000   
 & 6667 \\ 
\midrule
Power Effic. (KGPS/W) 
& 0.03  
& 0.21 
& 0.68 
& 2.36   
& \textcolor{\mycolor}{8.89} 
& \textcolor{\mycolor}{29.63} 
\\ 
\midrule
\end{tabular} }
   \begin{tablenotes}
    \footnotesize
        \item[1] Same JEDI-net architecture to U4 or J4.
    \normalsize
   \end{tablenotes}
\end{threeparttable}
\vspace{0.2cm}
\end{table}


%



\section{Related work}



Several studies have explored the use of GNNs for particle physics applications, such as jet tagging (identification)~\cite{moreno2020jedi}, charged particle tracking~\cite{ju2021performance}, and calorimeter energy measurements~\cite{qasim2019learning}. More can be found in a survey~\cite{thais2022graph}. 
To achieve low latency, FPGAs are utilized. The work in~\cite{elabd2021graph} extends the hls4ml~\cite{duarte2018fast} tool to translate GNNs into FPGA firmware automatically for charged particle tracking. GarNet~\cite{iiyama2021distance}, a GNN-based algorithm, is proposed for calorimeter energy regression.

Numerous studies also focus on general GNN accelerations~\cite{zhang2021boostgcn,lin2021gcn, geng2020awb, geng2021gcn, zhou2022model, tian2022g, garg2022understanding}. AWB-GCN~\cite{geng2020awb} is based on a column-wise-product architecture with runtime re-balancing for GCN acceleration. Their upgraded version I-GCN~\cite{geng2021gcn} presents Islandization, a new runtime graph restructuring algorithm, to improve data locality. BoostGCN~\cite{zhang2021boostgcn} presents a novel hardware-aware Partition-Centric Feature Aggregation (PCFA) scheme for pipelined GCNs.  
Lin~\textit{et al.}~\cite{lin2021gcn} introduce GCN acceleration using HLS and hardware-friendly optimizations. G-NMP~\cite{tian2022g} presents a Near-Memory Processing (NMP) solution for accelerating GNNs to handle the irregular memory access. Garg \textit{et al.}~\cite{garg2022understanding} explore various dataflow choices for sparse and dense GNNs on spatial accelerators. The work in~\cite{besta2022parallel} designs a taxonomy of parallelism in GNNs. Sohrabizadeh \textit{et al.}~\cite{sohrabizadeh2022streamgcn} present StreamGCN for accelerating GCN specialized for GNN streaming processing with small graphs.
Chen \textit{et al.}~\cite{chen2022regraph} introduce a heterogeneous pipeline architecture for GNNs on high bandwidth memory (HBM) enabled FPGAs. Abi-Karam \textit{et al.}~\cite{abi2022gengnn} propose GenGNN framework to deliver ultra-fast GNN inference and support a diverse set of GNN models. 
The results indicate that their designs achieve latency at the millisecond level. 
They also propose FlowGNN~\cite{sarkar2022flowgnn} which can flexibly support the majority of message-passing GNNs. Kang \textit{et al.}~\cite{kang2022grow} propose a GCN accelerator named GROW with Gustavson’s algorithm to architect a sparse-dense GEMM accelerator with row-wise product. Sun \textit{et al.}~\cite{sun2022multi} propose MultiGCN which balances network latency and network bandwidth for GCNs in multi-node systems. Yang \textit{et al.}~\cite{yang2022drgn} present a dynamically reconﬁgurable accelerator for GNNs named DRGN. EGCN~\cite{han2022egcn} is proposed using tiled matrix multiplication to reduce Off-Chip Memory Access. Furthermore, the I-GCN~\cite{geng2021gcn} adopts 4096 multiplier-accumulator (MAC) units running at 350MHz. Its peak performance is 2703 Giga Operations Per Second (GOPS). Each MAC is counted as 2 operations since it includes both multiplication and addition. BoostGCN~\cite{zhang2021boostgcn} has a peak performance of 1792 GOPS, running at 250MHz. GenGNN~\cite{abi2022gengnn} utilizes 1344 DSP blocks with 16-bit fixed-point data representation, resulting in a peak performance of 806 GOPS. Their later work, FlowGNN~\cite{sarkar2022flowgnn} has a peak performance of 1499 GOPS. Both of them run at 300MHz. Our design (\textbf{J4}), running at 200MHz, can not only provide a sub-microsecond latency but also an effective performance of 3025 GOPS which is higher than I-GCN, BoostGCN, GenGNN and FlowGNN. 
The custom MMMs, which do not contain any multiplications or additions, are excluded, with the exception of MMM3 that includes a small number of additions.
Finally, none of these previous designs target a sub-microsecond scenario. 
Although throughput at the application level can serve as a fair metric, the use of various applications for benchmarking makes it more complicated.



There are also previous studies about algorithm and hardware co-design for GNNs~\cite{zhou2022model, you2022gcod, zhang2021g, zhong2023cognn}. 
The work in~\cite{zhang2021g} presents a GNN and accelerator automatically co-search framework to maximize both task accuracy and acceleration efficiency. 
Zhou \textit{et al.}~\cite{zhou2022model} propose model-architecture co-design with a light-weight algorithm for temporal GNN inferences on FPGAs. 
You \textit{et al.}~\cite{you2022gcod} propose GCoD framwork, involving a two-pronged accelerator. 
Zhong~\textit{et al.}~\cite{zhong2023cognn} propose an algorithm-hardware co-design scheme with a reuse-aware sampling method to accelerate GNN inference in mini-batch scenarios. 
Some previous studies focus on accelerating GNN training~\cite{zeng2020graphact, su2021graph,lin2022hp, ogbogu2022accelerating}. GraphACT~\cite{zeng2020graphact} introduces an FPGA-based accelerator with a subgraph-based algorithm for Graph Convolutional Networks (GCNs) training. Su \textit{et al.}~\cite{su2021graph} present an efficient graph sampling accelerator on HBM enabled FPGAs for training GNNs. Lin \textit{et al.}~\cite{lin2022hp} propose HP-GNN which maps GNN training on the CPU-FPGA platform automatically. DietGNN~\cite{ogbogu2022accelerating}, a crossbar-aware pruning technique, is proposed to accelerate the training of large-scale GNNs. 

Most of the previous designs utilize a fixed hardware for all the layers of GNNs and process them sequentially. This is not efficient for GNN inference execution when targeting small graphs with requirements of ultra-low latency (such as $< 1\mu s$) and high throughput for scientific applications, such as particle identification.
This work focuses on layer-wise architecture for streaming processing of input graphs. We propose multiple novel optimizations to achieve high throughput and sub-microsecond level latency. 
These previous studies are orthogonal to our proposed approach and hardware architecture. Their techniques could be complementary to our approach, which could be extended in the future to achieve even lower latency.



\section{Conclusions and Future Work}

This paper presents a novel approach for minimizing the latency for processing GNNs on FPGAs, using JEDI-net algorithm as an end-to-end application. It involves optimizing the matrix operations and hardware pipeline to support next-generation low-latency collider trigger systems, the key to many fundamental physics experiments including particle identification.
Results show up to 9.0 times reduction in latency over the existing GPU-based JEDI-net implementation and up to 16.7 times reduction over state-of-the-art work. 
Future work includes exploring the use of new FPGA resources such as the AI Engines~\cite{xilinx_white} and the AI Tensor Blocks~\cite{langhammer2021stratix}, 
automating the proposed co-design approach using techniques such as meta-programming~\cite{vandebon2021enhancing, que2023metaml}, and incorporating the proposed techniques into the design and implementation of the data processing architecture for next-generation collider trigger systems. 

\section*{Acknowledgement}
The support of the United Kingdom EPSRC (grant numbers EP/V028251/1, EP/L016796/1, EP/N031768/1, EP/P010040/1, and EP/S030069/1), CERN, AMD and SRC are gratefully acknowledged.

\bibliographystyle{ACM-Reference-Format}
\bibliography{main-bib}

\end{document}